\documentclass[aps,prb,reprint,superscriptaddress]{revtex4-1}
\usepackage{bbold}
\usepackage{appendix}
\usepackage{H1}
\usepackage{amsmath}
\usepackage[pdftex,colorlinks=true]{hyperref}
\hypersetup{citecolor = blue}
\usepackage[normalem]{ulem}
\usepackage{amssymb}
\usepackage{amsthm}
\usepackage{amsfonts}
\usepackage{bbm}
\usepackage{array}
\usepackage{xcolor}
\usepackage{soul}

\begin{document}

\title{Obtaining highly excited eigenstates of the localized XX chain via DMRG-X }

\author{Trithep Devakul}
\affiliation{\mbox{Department of Physics, Princeton University, Princeton, NJ 08544, USA}}

\author{Vedika Khemani}
\affiliation{\mbox{Department of Physics, Harvard University, Cambridge, MA 02138, USA}}

\author{Frank Pollmann}
\affiliation{\mbox{Max-Planck-Institut f\"ur Physik komplexer Systeme, N\"othnitzer Str.\ 38, 01187 Dresden, Germany}}
\affiliation{Department of Physics, Technical University of Munich, 85748 Garching, Germany}

\author{David A. Huse}
\affiliation{\mbox{Department of Physics, Princeton University, Princeton, NJ 08544, USA}}

\author{S. L. Sondhi}
\affiliation{\mbox{Department of Physics, Princeton University, Princeton, NJ 08544, USA}}

\newcommand{\david}[1]{ { \color{red} {{#1}}}}
\newcommand{\vedika}[1]{ { \color{blue} {{#1}}}}

\begin{abstract}
We benchmark a variant of the recently introduced DMRG-X algorithm against exact results for the localized random field XX chain.  We find that the eigenstates obtained via DMRG-X exhibit a highly accurate l-bit description for system sizes much bigger than the direct, many body, exact diagonalization in the spin variables is able to access.  We take advantage of the underlying free fermion description of the XX model  to accurately test the strengths and limitations of this algorithm for large system sizes.
We discuss the theoretical constraints on the performance of the algorithm from the entanglement properties of the eigenstates, and its actual performance at different values of disorder. 
A small but significant improvement to the algorithm is also presented, which helps significantly with convergence.
%\david{Judging by our behavior since we obtained these results, the following sentence is false and should be removed:}  
%Altogether this study offers considerable encouragement for the use of the algorithm even for interacting many body localized spin models that do not possess an underlying free fermion description.
We find that at high entanglement, DMRG-X shows a bias towards eigenstates with low entanglement, but can be improved with increased bond dimension.
This result suggests that one must be careful when applying the algorithm for interacting many body localized spin models near a transition.
\end{abstract}

\maketitle

\section{Introduction}
The density matrix renormalization group (DMRG) has become one of the most successful numerical tools in the study of ground state properties of one-dimensional systems~\cite{White:1992,Schollwock2011}. DMRG is able to easily go to far larger sizes than accessible by methods such as exact diagonalization, and is even capable of working directly in the thermodynamic limit for translationally invariant systems~\cite{McCullochIDMRG}. The success of DMRG has been greatly
clarified by the advent of quantum information theoretic ideas such as the entanglement structure of quantum states and the language
of matrix product states (MPSs)~\cite{Schollwock2011,NorbertMPS}. The basic premise of DMRG is the relatively small entanglement present in the  ground states of most gapped local Hamiltonians. Specifically, such states obey an area law, that is, the bipartite entanglement entropy $S_E$ between two subsystems scales with the area of the boundary separating the regions~\cite{HastingsAreaLaw,AradAreaLaw}. As a consequence the states can be
efficiently represented with controllable error by MPSs which allow efficient computation of desired operator expectation values with controlled
errors. This {\it representability} is coupled in practice with {\it findability}---which is the existence of efficient algorithms, such as
the original DMRG algorithm, which find the desired representation in a time that scales polynomially ---indeed linearly---with system size.
While for the DMRG algorithm this scaling is an empirical fact about some unknown dynamical system\footnote{Although with possible exceptions, as discussed in Ref~\onlinecite{Eisert}}, there now also exists an algorithm that {\it provably} solves the problem in a time that scales polynomially in system size~\cite{dmrgproof}.
DMRG also works very well for critical points, where $S_E$ grows logarithmically with subsystem size which is modest enough to allow sufficient accuracy.

For thermalizing systems this cluster of ideas fails when addressed to highly excited eigenstates --- eigenstates with a finite energy density
corresponding to a nonzero temperature---due to the volume law entanglement that is generically present in these states. However in many body
localized (MBL) systems\cite{Anderson58,Basko06, PalHuse, OganesyanHuse,Nandkishore14,AltmanVosk} even highly excited eigenstates exhibit area law entanglement\cite{PalHuse,Bauer13} which leads to the possibility\cite{PekkerClark,ChandranSpectral}
that excited states can be constructed efficiently via DMRG-like algorithms in a manner analogous to ground states. Being able to evade the
size restrictions on many body exact diagonalization (MBED) in this fashion would be particularly useful in studying the eigenstate phase transition\cite{Huse13,PekkerHilbertGlass,EigenstateReview} from a volume-law obeying thermalizing phase at low disorder to an area-law obeying localized phase at strong disorder. 
While this MBL phase transition has been the subject of intense study numerically\cite{OganesyanMBL,KjallMBL,VoskMBL,PotterMBL,DevakulMBL,LuitzMBL,ZhangMBL,SerbynMBL,ZhangMBL2,KhemaniMBL}, there is still much that is not well understood. Even for the ``standard'' model of MBL\cite{PalHuse}, the random-field Heisenberg model, estimates of the critical disorder from numerical linked cluster expansion~\cite{Devakul15} predict a different value than finite size scaling estimates from exact diagonalization studies~\cite{PalHuse, Luitz15}. 
Moreover, a recent study\cite{KhemaniCP} has shown that the entanglement properties of eigenstates jump discontinuously at the MBL-to-thermal transition---even while other properties look continuous---and that the critical point has far less entanglement than previously assumed\cite{GroverCP}, thereby lending encouragement for the potential use of  DMRG-like methods all the way to the transition to the thermal phase. 
%These difficulties are believed to arise due to states which, at small lengthscales, appear localized but are actually thermalizing at much larger lengthscales than accessible by most numerical methods~\cite{nlc}.

Despite the good news on representability stemming from the area law, we note that the obstacles to constructing highly excited eigenstates of MBL systems are still formidable. One of these obstacles arises from the sheer smallness of gaps in the bulk of the many body spectrum which decrease exponentially with system size. Standard double precision arithmetic only allows a binary representation accurate to 52 bits which translates into an average spectral gap in a system of essentially the same number, say 50, spins. For systems larger than 50 spins we should expect to find exact spectral degeneracies which are artifacts and hence eigenvectors which are misleading superpositions of the underlying eigenstates. Indeed, this will start to happen even before size 50 for
selected pairs of levels which are closer than average. The second obstacle is the process of optimization in a random system. Almost all work on ground state DMRG has been done on undisordered systems and based on general experience with disordered systems there is reason to worry that optimization will not work as well in the more complicated landscapes the latter exhibit.
Potential obstacles here are the presence of Griffths regions and many-body resonances that may cause difficulty in finding an eigenstate.

With this caution injected, we note that the past year has nevertheless seen several studies that have tackled the problem of finding MPS representations for highly excited states. The first of these introduced an approach to finding approximations for {\it all} eigenstates of a MBL Hamiltonian at one go\cite{PollmannVUMPO} and
has been the object of more recent elaboration\cite{PalSimon}. The second, which is the basis of present work, is a variant of DMRG aimed to target highly excited eigenstates, called DMRG-X \cite{KhemaniDMRGX}. Rather than optimizing an MPS towards the ground state of the Hamiltonian, the DMRG-X algorithm iteratively optimizes the MPS towards an eigenstate of the Hamiltonian based on overlap with the state in the previous step. In a highly localized system, where eigenstates look very close to product states, DMRG-X was shown to converge very well starting from a product state as an initial state and reproduce the results of ED for accessible system sizes.  
In other works~\cite{YuPekkerClark,LimSheng,SerbynPowerlaw,Dante}, DMRG based techniques have been used to obtain excited eigenstates via energy (rather than overlap) targeting.

In this paper we return to the DMRG-X algorithm with a view to benchmarking its ability to accurately solve for eigenstates of large systems.
We now clarify what we mean by accurately and what we mean by large. One can usefully think of a single DMRG step for a many body system 
as a many body exact diagonalization (MBED) with soft boundaries so it amplifies the size one can study with a diagonalization subroutine of given power; this is what enables ground states of systems with 100 spins to be studied on a laptop where only 14 (say) could be studied by MBED. Correspondingly we wish to understand the amplification achieved by DMRG-X in the middle of the spectrum over MBED. To gauge this meaningfully
and even otherwise, we need to be able to assess the accuracy of the states obtained via DMRG-X.  For any system we can quantify the 
accuracy by computing the variance of the energy in the states that we obtain and comparing it to the level spacing; this criterion was used
in Ref.~\onlinecite{KhemaniDMRGX} at the limits of machine precision for the variance. However it is not known how to translate from a given accuracy of the variance
to the accuracy of various quantities of more physical interest, which limits the utility of this metric. 

To get around this limitation, in the present paper we study the the random field XX spin-1/2 chain which has an underlying free fermion representation. The free fermion character is not exploited by DMRG-X, which treats it as it does any interacting spin system. However the 
ability to exactly solve this model numerically  for large system sizes in the fermionic version allows us to check the results
of DMRG-X to larger system sizes than are accessible via MBED. Specifically, we make use a defining characteristic of localized phases, namely the existence of an emergent set of an extensive number of commuting $\mathbb{Z}_2$- valued local integrals of motion\cite{Huse14,Serbyn13cons} (often called ``l-bits''), and use the the free fermion representation to construct the exact l-bits for very large system sizes.  We then compute the expectation values of these l-bits in the DMRG-X obtained states, which allows us to directly assess how well these states approximate the exact eigenstates using a more physical (and economical) measure. 
 A further advantage of studying the XX chain is
that we are also able to study the representability of the eigenstates to large system sizes via computation of their entanglement entropies.  
Overall we find that DMRG-X achieves a roughly three-fold amplification over MBED when we require that all l-bits are accurate to at least 90 percent; most are vastly more accurate,
but, however, DMRG-X is biased towards less entangled states when entanglement becomes too high to represent.
While this particular demonstration is specific to the XX chain, we believe this analysis offers much encouragement that we can trust DMRG-X results even for interacting systems (away from the MBL phase transition). We note that similar optimism comes from the work of Serbyn et al\cite{SerbynPowerlaw} who examine the structure of the entanglement spectrum for numerically obtained eigenstates.

In the balance of the paper, we start with a quick recapitulation of the DMRG method in the language of MPSs (\secref{sec:mpsreview}). Following this, we review the basics of the random XX chain and the accuracy measure (\secref{sec:XX}), discuss representability where we also address the issue of rare Griffiths effects (\secref{sec:represent}), turn to findability where we find it important to modify the DMRG-X to use a hybrid metric for
the updates (\secref{sec:find}), show that the resulting algorithm is accurate enough to go beyond ED (\secref{sec:accuracy}) and end with some concluding remarks (\secref{sec:conclusion}). Appendices describe some technical details on the locality (\appref{sec:lbitlocality}) and representation (\appref{sec:lbitMPO}) of the  l-bit operators for the XX chain, and some unexpected wrinkles that crop up in studying the closely related problem of localization in quasiperiodic potentials (\appref{sec:qp}).

\begin{figure}[]
\includegraphics[width=\columnwidth]{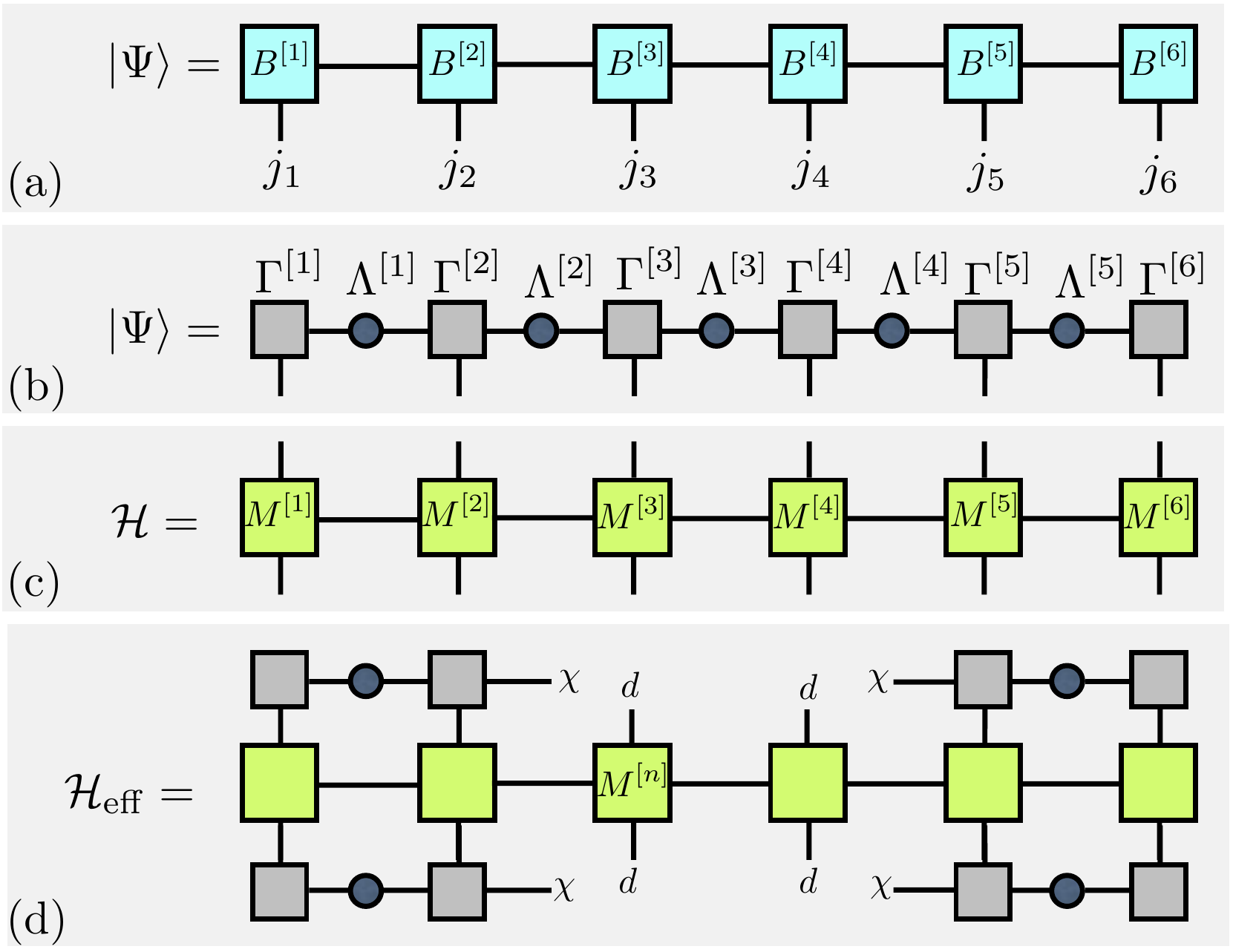} \\
\caption{Diagrammatic representation of (a) the state $|\Psi\rangle$ as an MPS, (b) $|\Psi\rangle$ as a canonical MPS, (c) the Hamiltonian $\mathcal{H}$ as an MPO,  and (d) the effective Hamiltonian $\mathcal{H}_\text{eff}$ in the variational basis.  }
 \label{fig:MPS}
\end{figure}

\section{Brief Review of MPS/MPO/DMRG formalism}
\label{sec:mpsreview}
In this section, we briefly recapitulate the standard DMRG algorithm\cite{White:1992} implemented in the language of matrix-product states (MPSs) \cite{Schollwock2011}. This section closely follows the supplementary material in Ref.~\onlinecite{KhemaniDMRGX}. A general quantum state $|\Psi\rangle$ for a one-dimensional system of $L$ sites can be written in the following matrix-product state (MPS) form:
\begin{equation}
	|\Psi \rangle = \sum_{j_1, \ldots, j_L} \sum_{0< \gamma_n <= \chi_n} B^{[1]j_1}_{\gamma_1} B^{[2]j_2}_{\gamma_1 \gamma_2} \ldots B^{[L]j_L}_{\gamma_{L-1}} | j_1, \ldots ,j_{L} \rangle.  \label{eq:mps}
\end{equation}
where $|j_n\rangle$ with $j_n=1,\dots,d$ is a basis of local states at site $n$ (for a spin 1/2 system, $d = 2$ and $|j_n\rangle = |\uparrow \rangle, |\downarrow\rangle$), and the  $B^{[n]}$ are rank three tensors (except on the first and last sites where they are rank two tensors). Figure \ref{fig:MPS}(a) shows a pictorial representation of an MPS. The enternal legs $j_n$ are the ``physical'' spin indices whereas the internal legs $\gamma_n$ are the virtual indices that are contracted. Each $B^{[n]j_n}$   is a $\chi_{n} \times \chi_{n+1}$ matrix (at the boundaries $\chi_1 = \chi_{L+1} = 1$) and each matrix product $\prod_i B^{[i]j_i}$ in~Eq.~(\ref{eq:mps}) produces a complex number which is the amplitude of $|\Psi \rangle$ on the basis state $|j_1 \cdots j_L \rangle$. 

The maximum dimension $\chi$ of the $\{B^{[n]}\}$ matrices is called the bond-dimension of the MPS and low entanglement states can be efficiently represented my MPSs of bond dimension $\chi \ll d^{L/2}$. The relationship between $\chi$ and the entanglement can be made more precise by considering the Schmidt decomposition of the state $|\Psi\rangle$. For a given bipartition of the system into left and right halves, a singular value decomposition can be used to rewrite $|\Psi\rangle  =  \sum_{\alpha} \Lambda_\alpha |\alpha \rangle_L |\alpha\rangle_R$
where the $|\alpha \rangle_{L/R}$ form orthonormal bases for the left and right halves respectively, and the entanglement entropy of the bipartition is defined through the Schmidt values $\Lambda_\alpha$ as $S_E = -\sum_\alpha |\Lambda_\alpha|^2 \ln |\Lambda_\alpha|^2$. Following a prescription by Vidal\cite{VidalCanonical}, it is possible to define a \emph{canonical form} (Fig.\ref{fig:MPS}(b)) for the MPS by rewriting each matrix $B^{[n]j_n}$ as a product of a $\chi_{n} \times \chi_{n+1}$ dimensional complex matrix $\Gamma^{[n]j_n}$ and a square diagonal matrix $\Lambda^{[n]}$ such that matrices $\Lambda^{[n]}$ matrices contain the non-zero Schmidt values for a bipartition between sites $n$ and $n+1$
\begin{align}
	|\Psi \rangle &= \sum_{j_1, \ldots, j_L}  \Gamma^{[1]j_1} \Lambda^{[1]} \Gamma^{[2]j_2} \Lambda^{[2]} \ldots \Lambda^{[L-1]}\Gamma^{[L]j_L} | j_1, \ldots ,j_{L} \rangle\nonumber \\
	 &\equiv \sum_{\alpha = 1}^{\chi_{n+1}}\Lambda^{[n]}_{\alpha \alpha} |\alpha_n\rangle_L |\alpha_n \rangle_R,  
	\label{eq:canonical}
\end{align}
and the states $|\alpha_n\rangle_L, |\alpha_n\rangle_R$
 define the orthonormal Schmidt states for the left and right halves of the bipartition respectively. 
This canonical form clearly relates the bond dimension $\chi$ to the number of Schmidt values contributing significantly to the entanglement entropy. 

Starting from an initial random MPS, the DMRG algorithm iteratively finds the ground state $|\psi_0\rangle$ by sweeping through the system and variationally optimizing the MPS matrices $B^{[n]j_n}$ on neighboring sites to locally minimize the energy $\langle \psi_0| \mathcal{H}| \psi_0\rangle$ (keeping the rest of the chain fixed). In the commonly used two-site update which simultaneously updates two sets of matrices $B^{[n]j_n}$ and $B^{[{n+1}]j_{n+1}}$, an effective Hamiltonian $\mathcal{H}_{\rm eff}$ is constructed by projecting $\mathcal{H}$ to a mixed $\chi_{n}\chi_{n+2}d^2$ dimensional basis. Here, the local basis states $|j_n\rangle|j_{n+1}\rangle$ represent the two updated sites, and the eigenstates of the reduced density matrix $|\chi_{n}\rangle_L|\chi_{n+2}\rangle_R$ compactly represent the environment to the left and right of the updated sites. The ground state of $\mathcal{H}_{\rm eff}$ is then found --- which is the optimal state for minimizing $\langle \psi_0| H| \psi	_0\rangle$  in this subspace ---and the matrices on sites $n, n+1$ are updated. The procedure is repeated for all sites until convergence is achieved.
The matrix-product operator (MPO) representation of $\mathcal{H}$, defined exactly analogously to Eq.~\ref{eq:mps} but now using 4-index tensors $M$, is shown in Fig.~\ref{fig:MPS}(c), and the effective Hamiltonian is depicted pictorially in Fig.~\ref{fig:MPS}(d). 

The only difference between the ground-state DMRG algorithm outlined above and the DMRG-X algorithm of Ref.~\onlinecite{KhemaniDMRGX} is in the update step. In the DMRG-X algorithm,  \emph{all} the $d^2 \chi^2$ eigenstates of $\mathcal{H}_\text{eff}$ are obtained instead of just its ground state, and the matrices $B^{[n]j_n}$ are updated  using the eigenstate of $\mathcal{H}_{\rm eff}$ with the maximum overlap with the previously found state in the iterative scheme.  As will be discussed in \secref{sec:represent}, we will present a modification to this algorithm at this step, where instead of simply choosing the eigenstate with maximal overlap, we also minimize the energy variance within a small subspace of states with high overlap.
The algorithm is initialized with an appropriate initial state which is perturbatively ``close'' to the true eigenstates of the MBL Hamiltonian, such as a product state.

\section{The XX Chain}
\label{sec:XX}
The model we study using DMRG-X is the spin-1/2 XX chain with random fields, 
\begin{equation}
    \mathcal{H} = -\sum_{i=1}^{L-1} \left( S^{x}_{i}S^{x}_{i+1} + S^y_i S^y_{i+1} \right) - \sum_{i=1}^{L} h_i S^z_i
    \label{}
\end{equation}
where $S^{x,y,z}_i$ are spin-1/2 operators on site $i$ and $h_i \in [-W,W]$ is chosen from a uniform distribution with width $2W$.
This model is Anderson localized for any nonzero $W$, and the localization length scales as $\xi \approx 25/W^2$ at low $W$~\cite{Kappus1981}, while at large $W$ the locator expansion yields $\xi \sim 1/\log W$. This model can be mapped via a Jordan-Wigner transformation onto a system of non-interacting fermions and solved exactly to very large system sizes.

Using the standard Jordan-Wigner substitutions for $S^{\pm}_i = S^x_i \pm i S^y_i$,
\begin{eqnarray}
    S^{+}_i =   {(-1)}^{\sum\limits_{j=1}^{i-1}c_j^\dagger c_j} c_i^\dagger,  \qquad     S^{-}_i =   {(-1)}^{\sum\limits_{j=1}^{i-1}c_j^\dagger c_j} c_i,
    \label{}
\end{eqnarray}
we arrive at a quadratic Hamiltonian in terms of the fermionic $c^\dagger_i, c_i$ operators
\begin{eqnarray}
    \mathcal{H} &=&  -\frac{1}{2}\sum_{i=1}^{L-1} \left( c^\dagger_i c_{i+1} + \text{h.c.} \right) - \sum_{i=1}^{L} h_i c^\dagger_i c_i\\
    &=& \sum_{i,j} c_i^\dagger H_{ij} c_j
    \label{}
\end{eqnarray}
up to a constant.
This single particle Hamiltonian $H_{ij}$ is only of dimension $L$, and can be diagonalized numerically, $H_{ij} = \sum_\alpha U_{i\alpha} \varepsilon_\alpha U^\dagger_{\alpha j}$, giving
\begin{eqnarray}
    \mathcal{H} &=&  \sum_{\alpha=1}^{L} \varepsilon_\alpha a_\alpha^\dagger a_\alpha
    \label{}
\end{eqnarray}
up to additive constants, 
where $a_\alpha^\dagger = \sum_{i} c^\dagger_i U_{i\alpha}$, and similarly $a_\alpha = \sum_{i} U^\dagger_{\alpha j} c_j$.  

In spin-language, we refer to the $\sigma^z_i = 2c^\dagger_i c_i - 1$ as \emph{physical}-bit (p-bit) operators and $\tau^z_\alpha = 2 a^\dagger_\alpha a_\alpha - 1$ as \emph{localized}-bit (l-bit) operators, in accordance with standard MBL nomenclature~\cite{Huse14}.
This non-interacting Hamiltonian then takes the simple form $\mathcal{H}=\sum_\alpha h_\alpha \tau^z_\alpha$, where $h_\alpha = \epsilon_\alpha/2$.
The $2^L$ eigenstates of $\mathcal{H}$ are then obtained by picking each l-bit to be $+1$ or $-1$, corresponding to a filled ($n_\alpha = 1$) or empty ($n_\alpha=0$) fermionic state respectively, and the many-body fermionic eigenstates are constructed as $\left| \left\{ n_\alpha  \right\} \right\rangle = \prod_\alpha a^\dagger_\alpha n_\alpha \left|0\right\rangle$.

It is interesting to note that the fermionic raising and lowering operators $a^\dagger_\alpha$ and $a_\alpha$ are inherently nonlocal in spin language due to the ${(-1)}^{\sum_i c^\dagger_i c_i}$ chain from the Jordan-Wigner transformation.
However, in MBL systems one tends to speak of local bosonic raising and lowering operators $\tau^\pm_\alpha$.
Bosonic raising and lowering operators can be constructed via a reverse Jordan-Wigner transormation from the $a^\dagger_\alpha$,$a_\alpha$ operators, but it is not \emph{a priory} clear that these operators are indeed local.\footnote{They are obviously local if one takes the definition of an MBL system to be the existence of a finite depth unitary transformation that diagonalizes the Hamiltonian.}
In \appref{sec:lbitlocality} we show that, indeed, these bosonic raising and lowering operators can be explicitly shown to be local given that the original l-bit operators are local.

Our primary interest in the l-bit operators is that they can be used to gauge the accuracy of a DMRG state. An exact eigenstate should have $|\langle \tau_\alpha^z \rangle|=1$ (in the absence of degeneracy).  Deviations from this tell us exactly where DMRG has failed to capture the state, and how badly it has done so.
The l-bit operators themselves can be efficiently expressed as matrix product operators (MPOs) using  an internal bond dimension of  only 4 (\appref{sec:lbitMPO}).

The other measure of accuracy is the total energy variance of the state, $\sigma_E^2 = |\langle H^2 \rangle - {\langle H \rangle}^2|$, which is also efficient to calculate for an MPS.
This should be closer to zero the closer a state is to an eigenstate of the Hamiltonian. Expressed in terms of the l-bit operators, $\sigma_E^2$ contains expectation values and correlators of these operators,
\begin{eqnarray}
    \sigma_E^{2} &=&  \sum_\alpha h_\alpha^2 (1 - {\langle \tau^z_\alpha \rangle }^2) \\
    &&+\sum_{\alpha \neq \beta} h_\alpha h_\beta \left( \langle \tau^z_\alpha \tau^z_\beta\rangle - \langle \tau^z_\alpha \rangle \langle \tau^z_\beta \rangle \right),
    \label{}
\end{eqnarray}
 and is thus a related but complementary measure of accuracy.

\section{Representability: $S_E$ and Griffiths Effects}
\label{sec:represent}
We now turn to the question of how efficiently the MB eigenstates of the XX model can be represented, which entails a study of the entanglement properties of these states. We find that rare Griffiths-like effects cause the typical and worst-case entanglement entropy across the bonds of the system to scale differently with $L$, which in turn affects the maximum bond-dimension needed to represent these states and the scaling of the computational time of the DMRG-X algorithm. While the algorithm always scales polynomially with $L$, in practice, the power of the polynomial can get quite large as disorder is lowered. 

The entanglement spectrum (ES) of a state across a cut determines how well that state can be represented as an MPS with a finite bond dimension $\chi$ on that cut.  
The ES in the MBL phase has been recently shown to decay as a power law~\cite{SerbynPowerlaw}.
However, the arguments of Ref~\onlinecite{SerbynPowerlaw} only apply to interacting MBL systems.\footnote{The argument in Ref~\cite{SerbynPowerlaw} relies on the l-bit operators flipping many spins within a radius $r$, but in a noninteracting model the l-bit operators only flip two.}
In the XX model, the spectrum of the single particle entanglement Hamiltonian $\mathcal{H}_E$ is found to have a constant density of states on average, similar to the clean system~\cite{Peschel}, which will lead to an entanglement spectrum decaying (on average) as 
\begin{equation}
 \lambda_n \sim \exp\left[-a {(\ln n)}^2 \right]   
    \label{eq:es}
\end{equation}
 for some constant $a$. This is one factor that differentiates our non-interacting model from an interacting model, making it easier to treat with DMRG-X\@. For now, we ignore the details of the estanglement spectrum and assume that a bond dimension of $\chi \sim e^{S_E}$ is needed to represent a state with entanglement
 entropy $S_E$ (which is equivalent to assuming a flat ES).

The entanglement entropy $S_E$ can be obtained exactly for a given many-body free-fermion eigenstate $|\psi\rangle = \left|\left\{ n_\alpha \right\}\right\rangle$ using the correlation matrix method~\cite{Peschel}.  
The idea is that the reduced density matrix for a partition $A$ is represented as a thermal density matrix for an ``entanglement Hamiltonian'' $\mathcal{H}_E$,
$\rho_\text{A} = \text{Tr}_{\bar{A}} \rho = e^{-\mathcal{H}_E}$, where $\mathcal{H}_E$ itself is a non-interacting Hamiltonian whose single particle energy eigenvalues can be found from the matrix of two-point  correlation functions of fermionic operators evaluated in the state $|\psi\rangle$.
The entanglement entropy $S_E = -\text{Tr} \rho_A \ln \rho_A$ can then be calculated knowing the eigenvalues of $\rho_A$.  
This knowledge can also be used to obtain the (many-body) eigenvalues of $\rho_\text{A}$, $\lambda_n$, which constitute the entanglement spectrum.\footnote{Finding the largest $k$ eigenvalues of $\rho_\text{A}$ involves finding the $k$ lowest total eigenenergies given all the single particle energies, which is a standard computational problem.}

\begin{figure*}
    \includegraphics[width=0.9\textwidth]{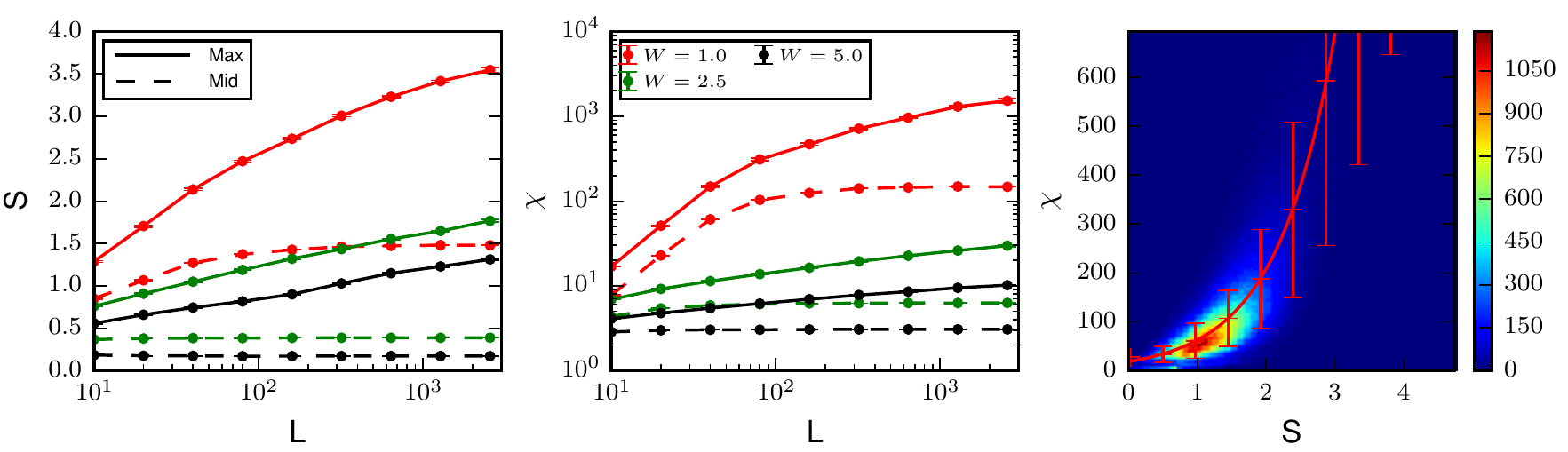}
    \caption{(left) The mean mid-bond and mean max entanglement entropy as a function of system size $L$.    
        These correspond to the entanglement on the middle bond and the maximum entanglement across all bonds within a sample, respectively, averaged over many samples.
        While the mean entanglement quickly saturates to an area-law value, the maximum entanglement within a sample continues increasing due to Griffiths regions. 
        (center) A similar plot of the mean mid and max $\chi$ as a function of system size, where $\chi$ is chosen such that on each cut, the total cutoff error is less than $10^{-5}$.  Similar to the entanglement entropy, the mean $\chi$ saturates to a constant, while the maximum continues increasing with system size.
        (right) A 2D histogram over all cuts and samples of bond entanglement and necessary bond dimension $\chi$ (assuming a naive truncation of the entanglement spectrum) to represent the state with total error\cite{Schollwock2011} less than $10^{-5}$ for systems size $L=320$ at $W=1.0$, demonstrating the average exponential dependence of $\chi$ on $S$.  }\label{fig:fig1}
\end{figure*}

Figure~\ref{fig:fig1}(left) shows the mean values of the mid-chain entanglement entropy and the maximum entanglement entropy within the whole chain.
The \emph{maximum} entanglement entropy $S_E^\text{max}$ for a given chain length is important as it will determine the maximum bond dimension $\chi$ needed to accurately capture that state as an MPS\@.
While the mid-chain entanglement entropy saturates to an area law, the maximum entanglement entropy continues to grow with $L$.

The growth of $S_E^\text{max}$ with $L$ can be understood as a type of Griffiths effect.
While there is no true delocalization in this model, a finite patch of length $\ell$ appears delocalized if the localization length within that patch exceeds $\ell$: $\xi \geq \ell$.  
Thus, the finite patch appears to have a critical disorder $W_c(\ell) = 5/\sqrt{\ell}$ (this follows from the expression for the localization length $\xi=25/W^2$).
The probability to get such a patch of length $\ell$ with every onsite field lying within  $W_c(\ell)$ is $P(\ell) = {(W_c(\ell)/W)}^{\ell}$.
Typically, a system of size $L$  will have $N(\ell) = L P(\ell)$ such patches, and equating $N(\ell)\approx1$ leads to $\ell \sim \log L / \log \log L$ at leading order.
%The largest $\ell$ that appears in that system will have $N(\ell)\approx1$, which in leading order leads to $\ell \sim \log L / \log \log L$.
This apparently delocalized patch will have volume law entanglement $S_E^\text{max} \sim \ell$, and therefore $S_E^\text{max} \sim \log L / \log \log L$, which explains the mostly logarithmic scaling with the slight downwards curvature in the $S_E^{\rm max}$ plot in Figure~\ref{fig:fig1}(left).
In particular, this analysis shows that the maximum bond dimension needed to represent a particular state which scales as $\log \chi^\text{max} \sim S_E^\text{max}$ will scale with $L$ (as has been previously observed~\cite{Bauer13,KhemaniDMRGX}), as opposed to approaching a constant as in the case of ground states of gapped systems.  
In \appref{sec:qp}, we discuss how a similar effect may arise even in some quasiperiodic systems, where there are no Griffiths effects.

This analysis also helps us deduce the scaling of the time needed to carry out one sweep of the DMRG-X. If we run the algorithm with variable bond dimension $\chi$ on each bond, the locus of the maximum computational time is determined by the tail of the 
distribution of entanglement entropies, with the shape of the tail determined by the same Griffiths-like effects that were just discussed.
If we take $S_E = \kappa \ell$ to be the entanglement of a thermal region of length $\ell$, assuming that the entanglement on each bond is independent, we have that 
\begin{eqnarray}
    P(S_E) &\sim \left(\frac{W_c(\ell(S_E))}{W}\right)^{\ell}  &\sim \exp\left[ -\frac{S_E}{\kappa} \ln \frac{W}{W_c(S_E/\kappa)} \right]
    \label{}
\end{eqnarray}
with $W_c(\ell)$ as defined earlier.
We therefore see that the tail of $P(S_E)$ will decay slightly faster than exponential, and decay faster for higher $W$.
However, the computational time required to deal with a cut of entanglement $S_E$  scales exponentially with $S_E$:
taking $\chi = e^{S_E}$, we know that the computational cost of diagonalizing the effective Hamiltonian in a DMRG step (a $d^2\chi^2 \times d^2\chi^2$ matrix) scales as $(\chi^2)^3 \sim e^{6S_E}$.
Thus, the mean computational time will be dominated by bonds with entanglement where $e^{6S_E}P(S_E)$ is maximized.
For low $W$, this implies that the computational time will be dominated by the largest entanglement across any bond in the sample $S_E^\text{max}$.
Thus, the growth of $S^\text{max}_E$ with $L$  leads to a complexity growing (almost) polynomially as $L^{b/\log\log L}$ with some constant $b$.
For higher $W$, as the tail decays more quickly, the dominating time shifts to an entanglement that appears more typically in the sample.
In this regime, the computational time appears to scale linearly with $L$.
In the case of a purely exponentially decaying tail in $P(S_E)$, the computation time experiences a Hagedorn-like transition at some $W$ where the dominating computational cost switches from being in the tail of the distribution to some finite value near the peak of the distribution.

Thus far we have assumed a flat ES in order to make the approximation $\chi = e^{S_E}$.
In reality, our ES decays on average quite quickly according to Eq~\ref{eq:es}, and the relationship between $\chi$ and $S_E$ depends on the exact structure of the ES.
However, $\chi$ should still scale exponentially with $S_E$ on average.
While the exact values of the coefficients of the exponential will depend on the decay of the ES and the error threshold, 
 the maximum bond dimension $\chi^\text{max}$ will still scale as predicted.
We confirm the dependence of $\chi$ on $S_E$ by sampling the entanglement spectra obtained from the correlation matrix~\cite{Peschel}.  
The largest values in the many-body entanglement spectra are obtained from the eigenvalues of the single particle entanglement Hamiltonian $\mathcal{H}_E$ via an iterative algorithm.
We find the largest $\chi$ eigenvalues such that the total sum of the discarded spectrum is less than $10^{-5}$.
Figure~\ref{fig:fig1}(right) shows a histogram of $\chi$ vs. $S_E$ at $W=1.0$, clearly demonstrating a rough exponential relation between the two as expected.  Notice that very large $\chi$, much larger than reachable, is needed to attain this level of accuracy for low $W$.
Figure~\ref{fig:fig1}(center) shows the max and mean $\chi$ as a function of $L$, demonstrating the expected scaling behaviour.  
At high $W$, $\chi$ becomes very small and the relationship between $S_E$ and $\chi$ is less apparent.

\section{Findability: Hybrid DMRG-X Algorithm}
\label{sec:find}
\begin{figure*}
    \includegraphics[width=1\textwidth]{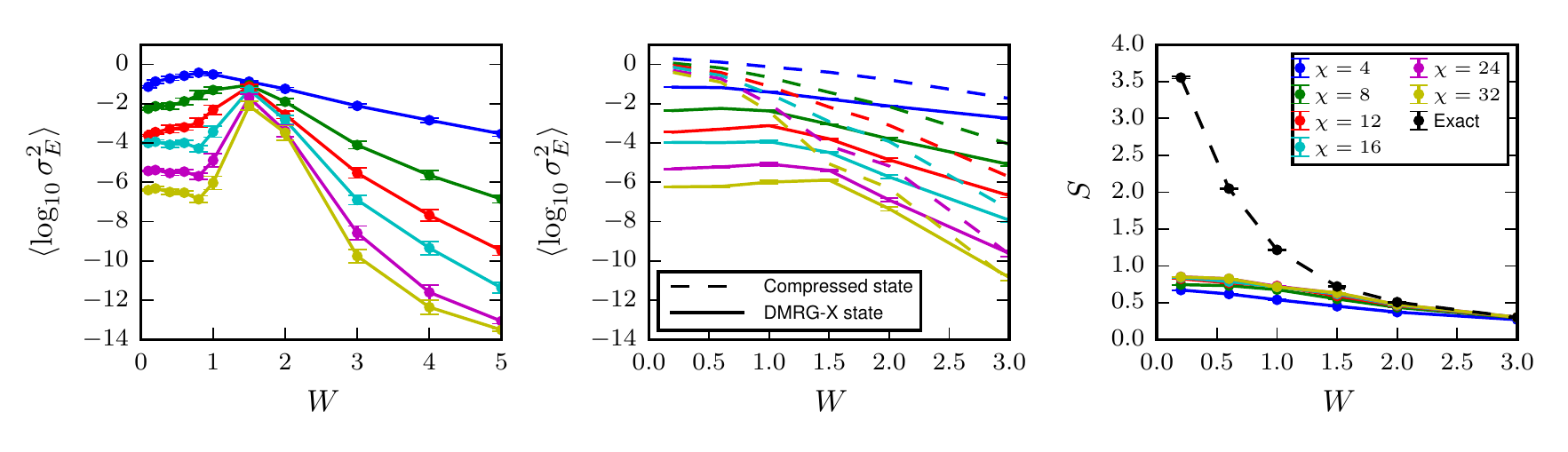}
    \caption{(left) Typical energy variance as a function of $W$ for a system of size $L=32$ for the original DMRG-X algorithm.  Error bars show the standard error of the mean.  DMRG-X works well at high $W$ where the eigenstates are highly localized, and does worse as $W$ is decreased.  There is a region near $W=1.5$ where the algorithm has the most difficulty converging onto an eigenstate.
    (center) Typical energy variance as a function of $W$ for a system of size $L=32$.  The hybrid variance minimization algorithm is used, greatly improving convergence within the difficult region.  The dashed lines show the variances of the compressed state obtained from the free fermion picture~\cite{WhiteCompression}, which DMRG-X always outperforms. 
    (right) The mean entanglement entropy obtained from DMRG-X using the hybrid algorithm for the $L=32$ system.  At low $W$, the mean obtained entanglement entropy is significantly lower than the mean values of the exact eigenstates, which indicates a bias towards low entanglement states.
    }\label{fig:fig2}
\end{figure*}

Having seen that the localized eigenstates of the XX model can be represented reasonably efficiently by MPS, we turn to the issue of finding such representations using the DMRG-X method. In performing DMRG-X for this relatively simple model, we found that the algorithm was having great difficulty converging for certain disorder realizations within a particular range of parameters.
The problem is not in the ability to represent such a state as an MPS, as better MPS representations can be constructed from the exact eigenstate, using an algorithm by S.R. White~\cite{WhiteCompression}.
Note that this algorithm is not the only way and may not be the optimal method to construct such MPSs, 
and in fact there has been much work on MPS representations for known states~\cite{Silvi,Katsura,Murg}.
White's algorithm iteratively applies pairwise rotations to an initial product state MPS, truncating the bond dimension as necessary, to construct a variational approximation---not necessarily the best one---to the exact eigenstate.
Given such an MPS as an initial state, performing a DMRG-X sweep will actually worsen the state when in the problematic parameter range ($W \approx 1.5$).
The typical energy variance $\sigma_E^2$ of the DMRG-X obtained eigenstates is shown for $L=32$ as a function of $W$ and $\chi$ in Figure~\ref{fig:fig2}(left).  
There is a particular region $1<W<2$ in which variance remains high and the algorithm fails to converge onto an eigenstate
A significant portion of samples within this region did not show a monotone decrease in variance with increasing $\chi$, meaning that the optimal state was never being reached.

A problem occurs when there is a near degeneracy between two many-body eigenstates that are not too many spin flips away.
If the dimension of our restricted subspace is not sufficient to accurately capture the Hamiltonian acting on these two states, they may appear to be much closer in energy than they actually are, resulting in a false resonance in $\mathcal{H}_\text{eff}$.
This happens as the subspace is optimized to accurately represent only the current state, not necessarily any others and therefore not the full Hamiltonian.
The result of this is that some of the eigenstates of $\mathcal{H}_\text{eff}$ are superpositions of the true eigenstate with other nearby energy states.
Thus, picking one of them based on maximal overlap does not move the overall state towards the correct answer.  
This is a problem in the findability. 

Here, we present a small modification that can correct for this type of error.  
Rather than simply picking the eigenstate with highest overlap, we look at the $n$ largest overlap eigenstates. 
The variance can then be optimized within this $n$-dimensional subspace, via a gradient method starting with the current state as a seed.
Thus, variance and overlap are both taken into account in choosing the next state, which is closer to the true eigenstate than any eigenstate of $\mathcal{H}_\text{eff}$.

The variance calculation can be done efficiently in MPS language by keeping track of $\mathcal{H}^2_\text{eff}$ in addition to just $\mathcal{H}_\text{eff}$.
Note that $\mathcal{H}^2_\text{eff}$ here is $\mathcal{H}^2$ in the effective subspace, not simply $\mathcal{H}_\text{eff}$ squared.
%These matrices are shown in MPS notation in Figure~\ref{fig:hdrawing}.
Then, the variance $\sigma_E^2 = \left| \langle\mathcal{H}_\text{eff}^2 \rangle - {\langle \mathcal{H}_\text{eff} \rangle}^2 \right|$ can be optimized using a gradient based optimization algorithm such as the BFGS algorithm. 
These matrices can be further reduced to $n$ by $n$ matrices by only looking in the subspace of the $n$ states with highest overlap with the previous state.
An important implementation detail is that we optimize $\log \sigma^2_E$ rather than simply $\sigma^2_E$, as we are concerned with very small values of the variance.

\begin{figure*}
    \includegraphics{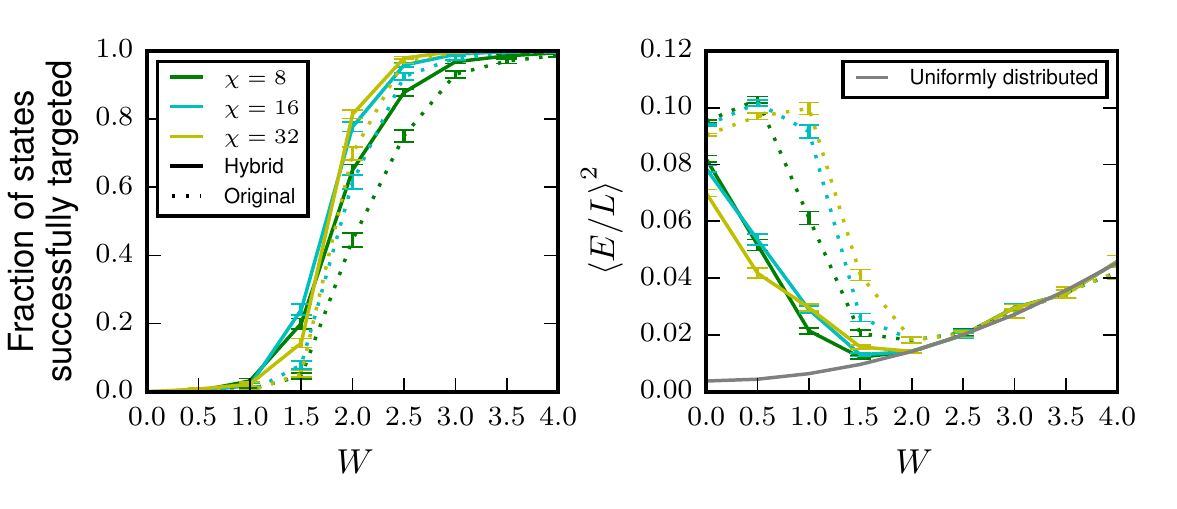}
    \caption{Calculations done starting from an MPS representation of a randomly chosen eigenstate, constructed as described in the text.  (left) The fraction of DMRG-X runs that eventually converge upon the initially chosen eigenstate.  At high $W$, one can essentially target the final state exactly, while at low $W$ the algorithm always finds another state to converge to.
        (right) The final DMRG-X energy density $\langle E / L\rangle$ squared and then averaged over samples.
        %\hl{\textbf{What is the best notation for this quantity?}}
        If we always converged upon the desired eigenstate, the energy would be uniformly distributed and this value would be equivalent to the average value of $2^{-L}\text{Tr}\mathcal{H}^2/L^2$, which is also shown in the plot.
        At low $W$, the bias towards lower entanglement entropy biases states towards the edges of the spectrum, which increases the obtained $\langle E/L\rangle^2$ above the uniform value.
        Both original and hybrid algorithm are subject to this bias, although to different degrees.
    }\label{fig:drift}
\end{figure*}

This hybrid algorithm works best when the MPS is already close to an eigenstate, as  the overlap then provides a useful measure of which states to optimize over.  
It is therefore often helpful to run a few sweeps with $n=1$ (the original algorithm) before increasing $n$.  
Rather than keeping $n$ fixed, it is also possible to vary $n$.  
If we let $\left|\phi_i\right\rangle$ be the eigenstates of the effective Hamiltonian ordered from highest to lowest overlap with the previous state $\left|\psi\right\rangle$, then we can choose $n$ to be the smallest integer satisfying
\begin{equation}
    1-\sum_{i=1}^{n} {\left| \langle \phi_i | \psi \rangle \right|}^2 < \epsilon
    \label{}
\end{equation}
where $\epsilon$ is chosen to be a small number ($\epsilon=10^{-10}$ is used in our calculations).

Figure~\ref{fig:fig2}(center) shows that the application of this hybrid algorithm greatly improves the variance near the troublesome region. 
The peak in the variance of the original algorithm arose due to convergence issues, which have been clearly resolved with the hybrid algorithm.
Also shown are the variances obtained from the MPS constructed using White's algorithm~\cite{WhiteCompression}.\footnote{Note that it is not that the algorithm of Ref~\cite{WhiteCompression} is failing at low $W$.  Similar curves were obtained for smaller $L$ accessible by ED, and a standard MPS compression of the ED state yielded similar results.}
Those variances do what one might expect from a direct truncation in the entanglement spectrum, in that they perform poorly at low $W$ where there is high entanglement and monotonically do worse.

Surprisingly DMRG-X appears to be performing well even at low $W$, where it is expected to fail.
Exploring further, Figure~\ref{fig:fig2} shows the mean entanglement entropy of the DMRG-X states in comparison the mean of the exact eigenstates.  
We see that the state to which DMRG-X converges towards is significantly lower in entanglement than the true exact eigenstates.
The reason for this is that DMRG-X is biased towards states which it can represent well with limited $\chi$.  
That is, when the available $\chi$ is small, the algorithm is biased towards states with low entanglement entropy, which lie away from the center of the spectrum in energy.
At very low $W$, the only representable states for small $\chi$ are near the edges of the spectrum, which DMRG-X finds.

The primary reason this occurs is that the algorithm allows the state to ``drift'' with each iteraction.  
This is in contrast with, say, the naive energy-targeting algorithm in which the eigenstate of the effective Hamiltonian closest to a specified energy is picked at each step (which fails to converge due to small many-body level spacings~\cite{KhemaniDMRGX}).  
The closely related algorithm ES-DMRG~\cite{YuPekkerClark}, which works by selecting the eigenstate with the energy closest to the energy of the previous state, is also subject to this ``drift'' towards low entanglement states at low $W$.
%Note that the hybrid DMRG-X algorithm is able to converge in significantly fewer sweeps than either of the previous algorithms 
%Thinking of the iterations of DMRG-X as that of some unknown dynamical system, what is happening is that eigenstates that cannot be represented well enough with the given bond dimension are no longer stable fixed points for the system.

In Fig~\ref{fig:drift} we show hybrid DMRG-X calculations, where an approximated eigenstate MPS (constructed as described earlier) has been given as seed.
In contrast to previous calculations where the bond dimension $\chi$ is ramped up slowly, in these calculations each choice of $\chi$ represents an independent calculation.
As seen in Fig~\ref{fig:drift}(left), at high $W$, DMRG-X (both hybrid and original) is able to converge upon the chosen eigenstate very quickly.
However, as $W$ is lowered and the entanglement in these eigenstates increases, neither are able to converge to the desired target.
Instead, a compromise is made and the algorithm converges upon a different eigenstate with lower entanglement away from the center of the spectrum.
Increasing the bond dimension $\chi$ helps extend this region of convergence.
As seen in Fig~\ref{fig:drift}(right), the hybrid algorithm suppresses this drift towards extremal energies somewhat, but is still subject to it.

\section{Accuracy of DMRG-X}
\label{sec:accuracy}
We now turn to our goal, which is to evaluate the performance of DMRG-X on system sizes larger than accessible by many-body ED. To this end
we use a modest bond dimension of $\chi=32$, corresponding to a $2^2\chi^2=2^{12}$ dimensional effective Hamiltonian in the worst case.
Thus the computational effort at the diagonalization step of DMRG-X is equivalent to the exact diagonalization of the full Hamiltonian of an $L=12$ system. With this restriction we examine the accuracy of the algorithm on system sizes much bigger than $L=12$ via the energy variance
and, more revealingly, l-bit accuracy.

\begin{figure*}
    \includegraphics[width=0.8\textwidth]{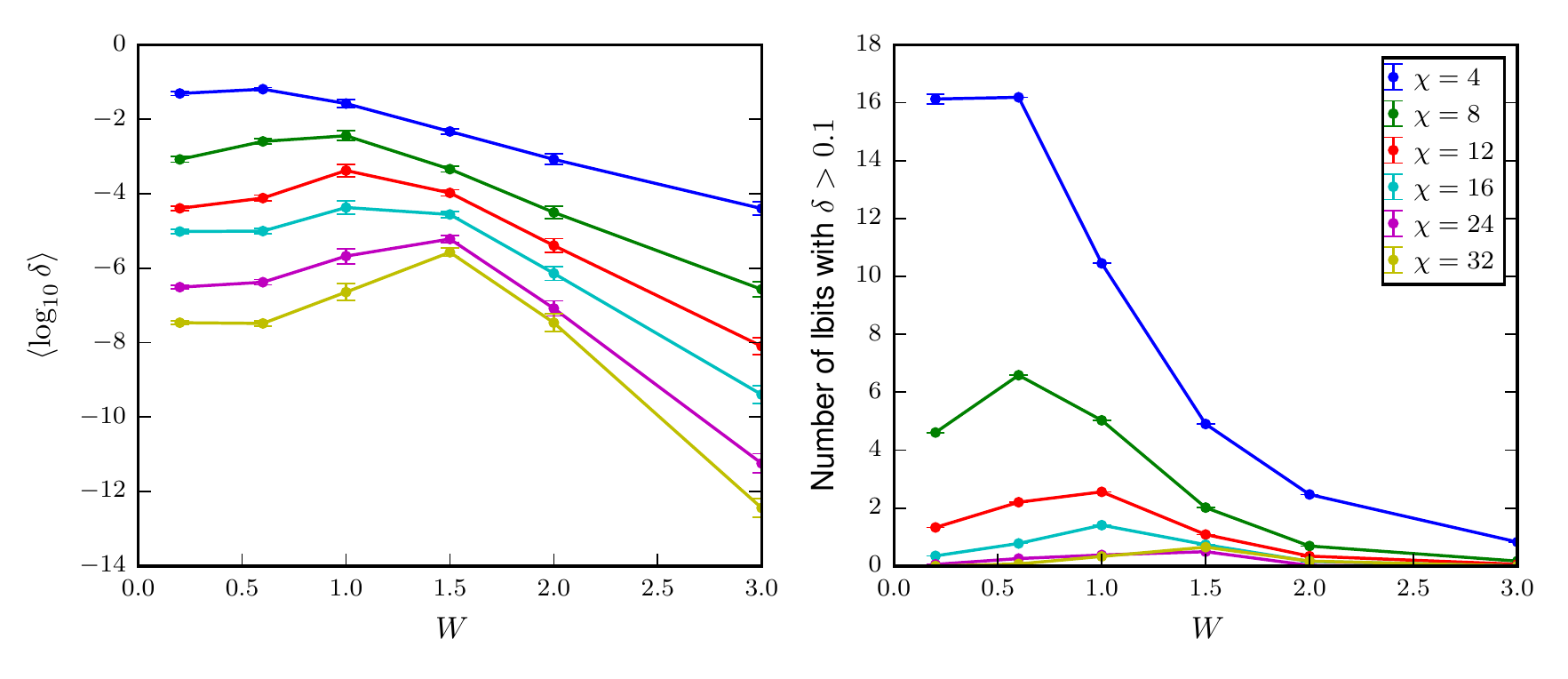}
    \caption{(left) Typical value of $\delta$ for l-bits in a system of size $L=32$, averaged over roughly 1000 disorder samples, with one eigenstate per sample.
(right)Average number of l-bits with $\delta > 0.1$ in a system of size $L=32$.
    }\label{fig:fig3}
\end{figure*}

%We begin by studying $L=32$. 
%We have seen in Fig~\ref{fig:fig2}(center) earlier that for $L=32$, even in the low $W$ regime where the localization length is larger than $L$ (and significantly larger than the many-body ED length for $\chi=32$), the algorithm is still able to converge to states to variances of $\sigma_E^2 \approx 10^{-6}$.
%Given the amount of entanglement in the low $W$ region, one would have naively expected DMRG-X to perform very poorly at low $W$.
%The ``flattening out'' of the energy variance at low $W$ and the lowered entanglement entropy is therefore DMRG-X performing better than expected --- it is able to find a state significantly lower in entanglement while also maintaining a low energy variance.
%In this case, one may worry about how close the state is to the actual eigenstates.  

We begin by studying $L=32$. 
To better understand the quality of the state, Fig~\ref{fig:fig3}(left) shows the typical deviation of $\langle \tau^z_\alpha \rangle$ from $\pm1$, defined by $\delta = 1-\left|\langle\tau^z_\alpha\rangle\right|$.
At $\chi=32$, the typical l-bit expectation value is accurate up to roughly $10^{-6}$ at its worst.
Figure~\ref{fig:fig3}(right) shows the average number of badly captured l-bits in a sample, which we define to be those with $\delta > 0.1$.
Indeed, with high enough bond dimension DMRG-X is able to obtain most eigenstates with no bad l-bits, with the mean number of bad l-bits falling below 1. This is our central demonstration in this paper---that DMRG-X can obtain all of the l-bits correctly for system sizes bigger than those treatable by the ED routine utilized by the code. Evidently, if we were to use the actual maximum size treatable by ED ($L>12$) we would expect to get to even bigger actual system sizes ($L>32$) and still expect to get all of the l-bits correctly by our metric. 
%\vedika{Minor point, but if it's easy, can we replace this with data for L = 40? It was mildly confusing reading the paragraph above with the same value for $L$ and $\chi$. I had to read twice because I confused the two} 

\begin{figure*}
    \includegraphics[width=0.8\textwidth]{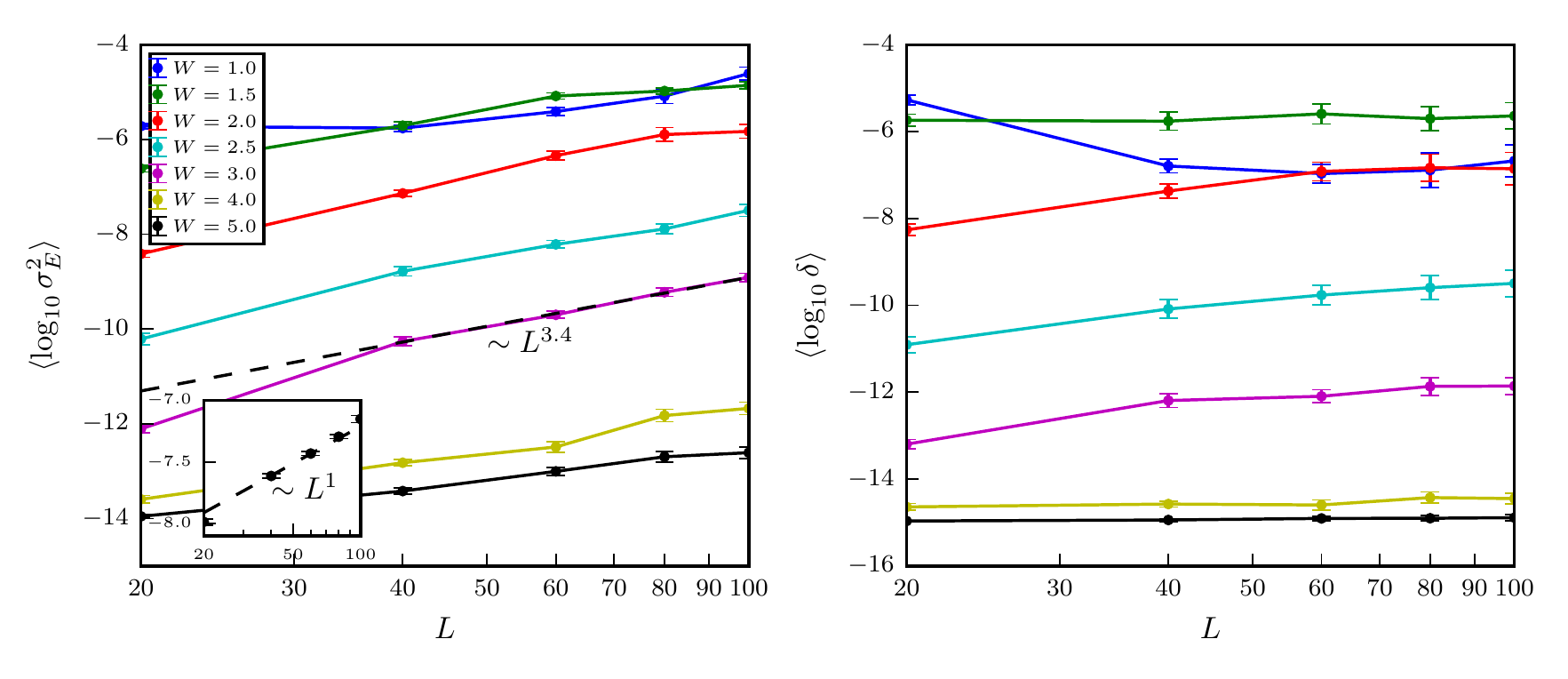}
    \caption{(left) Typical variance scaling with system size $L$ at fixed bond dimension $\chi=32$.  
        The inset shows the same plot for $W=5$ with fixed error density, truncating the Schmidt spectrum at a variable bond dimension such that the total cut-off error is less than $10^{-10}$, which shows a linear scaling as expected.
(right) Typical $\delta$ of l-bits scaling with system size $L$ at fixed bond dimension $\chi=32$.
        For high $W$, the typical l-bit is captured to machine precision, flattening out at $\langle \log_{10}\delta \rangle \approx 10^{-15}$
    }\label{fig:fig4}
\end{figure*}

We now turn to the scaling with $L$ while keeping $\chi=32$ fixed. 
If the MPS exhibited a constant error density throughout, one would expect the energy variance to scale linearly with $L$.
Figure~\ref{fig:fig4}(left) instead shows the variance scaling with $L$, which appears to be faster than linear.
The typical l-bit errors, shown in Figure~\ref{fig:fig4}(right), are also very small and slowly increasing with $L$.
The reason behind the super-linear growth in variance is that our states are in fact \emph{not} of constant error density, as we are simply truncating each bond at $\chi$ 
singular values.
To construct a state of constant error density, one would have to consider a variable $\chi$ on each bond such that the total discarded Schmidt weight is below an error threshold.
In such a scheme one would still need to define a maximum allowable $\chi$, $\chi_\text{max}$, such that the computation completes within a reasonable timeframe.
One can then either discard samples that would exceed this limit (which avoids the issue of potential Griffiths regions) or truncate the bonds at $\chi_\text{max}$ (which no longer produces constant error density states).
Our fixed-$\chi$ choice corresponds to the latter with a zero error threshold.

\section{Concluding Remarks}
\label{sec:conclusion}
We have tested the performance of the DMRG-X algorithm for finding highly excited eigenstates of the disordered XX model.  
With some small improvements to the algorithm, we were able to go significantly beyond the range of system sizes accessible to exact diagonalization in the many-body Hilbert space.
Most of the eigenstates are able to be obtained to high accuracy with only $\chi=32$, using the l-bit expectation value as a metric.
Thus, we have successfully shown that DMRG-X is able to go well beyond exact diagonalization in the well localized regime, converging to eigenstates quickly with increasing $\chi$.
For more delocalized states, DMRG-X makes a compromise to improve accuracy and convergence by biasing itself towards lower entanglement states.

There are still many ways of improving the capabilities of the algorithm.
Rather than exact diagonalization of the effective Hamiltonian, shift-invert Lanczos can be used to obtain all the eigenstates within an energy window.
The states with high overlap with the original state typically have very similar energy, so Lanczos is also able to find a relevant subspace of $n$ states over which to optimize variance. This theoretically allows a great increase in the range of accessible $\chi$.

The natural next step is to consider application of DMRG-X to interacting systems which exhibit MBL.
%However, the low entanglement bias of DMRG-X cast.s doubts on its ability to accurately capture this transition, where Griffiths regions with high entanglement play an important role.
Near the transition, Griffiths regions with high entanglement play an important role, thus one must be careful when applying DMRG-X near the transition.
This points out the need for larger bond dimensions which may be obtained as previously mentioned. 
%With increased confidence in the algorithm in hand we can now return to studying the interacting problem. In addition to the transition to the ergodic phase in the model with interactions added to the XX chain, another potential application is to the phase transition between two localized phases, for example the spin glass to paramagnetic phase transition in the random transverse field Ising model.

\begin{acknowledgments}
%The authors would like to thank...for useful discussions. 
This research was supported by the Harvard Society of Fellows (VK), the
DFG (Deutsche Forschungsgemeinschaft) Research Unit FOR 1807 through grants numbers PO 1370/2-1 (FP), the NSF-DMR via grant number
1311781 (SLS) and the Alexander von Humboldt Foundation via a Humboldt Award (SLS). 
\end{acknowledgments}

\bibliography{global}

\begin{thebibliography}{55}%
\makeatletter
\providecommand \@ifxundefined [1]{%
 \@ifx{#1\undefined}
}%
\providecommand \@ifnum [1]{%
 \ifnum #1\expandafter \@firstoftwo
 \else \expandafter \@secondoftwo
 \fi
}%
\providecommand \@ifx [1]{%
 \ifx #1\expandafter \@firstoftwo
 \else \expandafter \@secondoftwo
 \fi
}%
\providecommand \natexlab [1]{#1}%
\providecommand \enquote  [1]{``#1''}%
\providecommand \bibnamefont  [1]{#1}%
\providecommand \bibfnamefont [1]{#1}%
\providecommand \citenamefont [1]{#1}%
\providecommand \href@noop [0]{\@secondoftwo}%
\providecommand \href [0]{\begingroup \@sanitize@url \@href}%
\providecommand \@href[1]{\@@startlink{#1}\@@href}%
\providecommand \@@href[1]{\endgroup#1\@@endlink}%
\providecommand \@sanitize@url [0]{\catcode `\\12\catcode `\$12\catcode
  `\&12\catcode `\#12\catcode `\^12\catcode `\_12\catcode `\%12\relax}%
\providecommand \@@startlink[1]{}%
\providecommand \@@endlink[0]{}%
\providecommand \url  [0]{\begingroup\@sanitize@url \@url }%
\providecommand \@url [1]{\endgroup\@href {#1}{\urlprefix }}%
\providecommand \urlprefix  [0]{URL }%
\providecommand \Eprint [0]{\href }%
\providecommand \doibase [0]{http://dx.doi.org/}%
\providecommand \selectlanguage [0]{\@gobble}%
\providecommand \bibinfo  [0]{\@secondoftwo}%
\providecommand \bibfield  [0]{\@secondoftwo}%
\providecommand \translation [1]{[#1]}%
\providecommand \BibitemOpen [0]{}%
\providecommand \bibitemStop [0]{}%
\providecommand \bibitemNoStop [0]{.\EOS\space}%
\providecommand \EOS [0]{\spacefactor3000\relax}%
\providecommand \BibitemShut  [1]{\csname bibitem#1\endcsname}%
\let\auto@bib@innerbib\@empty
%</preamble>
\bibitem [{\citenamefont {White}(1992)}]{White:1992}%
  \BibitemOpen
  \bibfield  {author} {\bibinfo {author} {\bibfnamefont {S.~R.}\ \bibnamefont
  {White}},\ }\href {\doibase 10.1103/PhysRevLett.69.2863} {\bibfield
  {journal} {\bibinfo  {journal} {Phys. Rev. Lett.}\ }\textbf {\bibinfo
  {volume} {69}},\ \bibinfo {pages} {2863} (\bibinfo {year}
  {1992})}\BibitemShut {NoStop}%
\bibitem [{\citenamefont {Schollwock}(2011)}]{Schollwock2011}%
  \BibitemOpen
  \bibfield  {author} {\bibinfo {author} {\bibfnamefont {U.}~\bibnamefont
  {Schollwock}},\ }\href {\doibase http://dx.doi.org/10.1016/j.aop.2010.09.012}
  {\bibfield  {journal} {\bibinfo  {journal} {Annals of Physics}\ }\textbf
  {\bibinfo {volume} {326}},\ \bibinfo {pages} {96 } (\bibinfo {year}
  {2011})}\BibitemShut {NoStop}%
\bibitem [{\citenamefont {McCulloch}(2008)}]{McCullochIDMRG}%
  \BibitemOpen
  \bibfield  {author} {\bibinfo {author} {\bibfnamefont {I.~P.}\ \bibnamefont
  {McCulloch}},\ }\href@noop {} {\bibfield  {journal} {\bibinfo  {journal}
  {arXiv:0804.2509}\ } (\bibinfo {year} {2008})}\BibitemShut {NoStop}%
\bibitem [{\citenamefont {Schuch}\ \emph {et~al.}(2008)\citenamefont {Schuch},
  \citenamefont {Wolf}, \citenamefont {Verstraete},\ and\ \citenamefont
  {Cirac}}]{NorbertMPS}%
  \BibitemOpen
  \bibfield  {author} {\bibinfo {author} {\bibfnamefont {N.}~\bibnamefont
  {Schuch}}, \bibinfo {author} {\bibfnamefont {M.~M.}\ \bibnamefont {Wolf}},
  \bibinfo {author} {\bibfnamefont {F.}~\bibnamefont {Verstraete}}, \ and\
  \bibinfo {author} {\bibfnamefont {J.~I.}\ \bibnamefont {Cirac}},\ }\href
  {\doibase 10.1103/PhysRevLett.100.030504} {\bibfield  {journal} {\bibinfo
  {journal} {Phys. Rev. Lett.}\ }\textbf {\bibinfo {volume} {100}},\ \bibinfo
  {pages} {030504} (\bibinfo {year} {2008})}\BibitemShut {NoStop}%
\bibitem [{\citenamefont {Hastings}(2007)}]{HastingsAreaLaw}%
  \BibitemOpen
  \bibfield  {author} {\bibinfo {author} {\bibfnamefont {M.~B.}\ \bibnamefont
  {Hastings}},\ }\href {http://stacks.iop.org/1742-5468/2007/i=08/a=P08024}
  {\bibfield  {journal} {\bibinfo  {journal} {Journal of Statistical Mechanics:
  Theory and Experiment}\ }\textbf {\bibinfo {volume} {2007}},\ \bibinfo
  {pages} {P08024} (\bibinfo {year} {2007})}\BibitemShut {NoStop}%
\bibitem [{\citenamefont {I.~Arad}\ and\ \citenamefont
  {Vazirani}(2013)}]{AradAreaLaw}%
  \BibitemOpen
  \bibfield  {author} {\bibinfo {author} {\bibfnamefont {Z.~L.}\ \bibnamefont
  {I.~Arad}, \bibfnamefont {A.~Kitaev}}\ and\ \bibinfo {author} {\bibfnamefont
  {U.}~\bibnamefont {Vazirani}},\ }\href@noop {} {\bibfield  {journal}
  {\bibinfo  {journal} {arXiv:1301.1162}\ } (\bibinfo {year}
  {2013})}\BibitemShut {NoStop}%
\bibitem [{Note1()}]{Note1}%
  \BibitemOpen
  \bibinfo {note} {Although with possible exceptions, as discussed in
  Ref~\protect \rev@citealpnum {Eisert}}\BibitemShut {NoStop}%
\bibitem [{\citenamefont {{Landau Zeph}}\ \emph {et~al.}(2015)\citenamefont
  {{Landau Zeph}}, \citenamefont {{Vazirani Umesh}},\ and\ \citenamefont
  {{Vidick Thomas}}}]{dmrgproof}%
  \BibitemOpen
  \bibfield  {author} {\bibinfo {author} {\bibnamefont {{Landau Zeph}}},
  \bibinfo {author} {\bibnamefont {{Vazirani Umesh}}}, \ and\ \bibinfo {author}
  {\bibnamefont {{Vidick Thomas}}},\ }\href {\doibase
  http://dx.doi.org/10.1038/nphys3345 10.1038/nphys3345} {\bibfield  {journal}
  {\bibinfo  {journal} {Nat Phys}\ }\textbf {\bibinfo {volume} {11}},\ \bibinfo
  {pages} {566} (\bibinfo {year} {2015})}\BibitemShut {NoStop}%
\bibitem [{\citenamefont {Anderson}(1958)}]{Anderson58}%
  \BibitemOpen
  \bibfield  {author} {\bibinfo {author} {\bibfnamefont {P.~W.}\ \bibnamefont
  {Anderson}},\ }\href {\doibase 10.1103/PhysRev.109.1492} {\bibfield
  {journal} {\bibinfo  {journal} {Phys. Rev.}\ }\textbf {\bibinfo {volume}
  {109}},\ \bibinfo {pages} {1492} (\bibinfo {year} {1958})}\BibitemShut
  {NoStop}%
\bibitem [{\citenamefont {Basko}\ \emph {et~al.}(2006)\citenamefont {Basko},
  \citenamefont {Aleiner},\ and\ \citenamefont {Altshuler}}]{Basko06}%
  \BibitemOpen
  \bibfield  {author} {\bibinfo {author} {\bibfnamefont {D.~M.}\ \bibnamefont
  {Basko}}, \bibinfo {author} {\bibfnamefont {I.~L.}\ \bibnamefont {Aleiner}},
  \ and\ \bibinfo {author} {\bibfnamefont {B.~L.}\ \bibnamefont {Altshuler}},\
  }\href {\doibase 10.1016/j.aop.2005.11.014} {\bibfield  {journal} {\bibinfo
  {journal} {Annals of Physics}\ }\textbf {\bibinfo {volume} {321}},\ \bibinfo
  {pages} {1126} (\bibinfo {year} {2006})}\BibitemShut {NoStop}%
\bibitem [{\citenamefont {Pal}\ and\ \citenamefont {Huse}(2010)}]{PalHuse}%
  \BibitemOpen
  \bibfield  {author} {\bibinfo {author} {\bibfnamefont {A.}~\bibnamefont
  {Pal}}\ and\ \bibinfo {author} {\bibfnamefont {D.~A.}\ \bibnamefont {Huse}},\
  }\href {\doibase 10.1103/PhysRevB.82.174411} {\bibfield  {journal} {\bibinfo
  {journal} {Phys. Rev. B}\ }\textbf {\bibinfo {volume} {82}},\ \bibinfo
  {pages} {174411} (\bibinfo {year} {2010})}\BibitemShut {NoStop}%
\bibitem [{\citenamefont {Oganesyan}\ and\ \citenamefont
  {Huse}(2007{\natexlab{a}})}]{OganesyanHuse}%
  \BibitemOpen
  \bibfield  {author} {\bibinfo {author} {\bibfnamefont {V.}~\bibnamefont
  {Oganesyan}}\ and\ \bibinfo {author} {\bibfnamefont {D.~A.}\ \bibnamefont
  {Huse}},\ }\href {\doibase 10.1103/PhysRevB.75.155111} {\bibfield  {journal}
  {\bibinfo  {journal} {Phys. Rev. B}\ }\textbf {\bibinfo {volume} {75}},\
  \bibinfo {pages} {155111} (\bibinfo {year} {2007}{\natexlab{a}})}\BibitemShut
  {NoStop}%
\bibitem [{\citenamefont {Nandkishore}\ and\ \citenamefont
  {Huse}(2015)}]{Nandkishore14}%
  \BibitemOpen
  \bibfield  {author} {\bibinfo {author} {\bibfnamefont {R.}~\bibnamefont
  {Nandkishore}}\ and\ \bibinfo {author} {\bibfnamefont {D.~A.}\ \bibnamefont
  {Huse}},\ }\href {\doibase 10.1146/annurev-conmatphys-031214-014726}
  {\bibfield  {journal} {\bibinfo  {journal} {Annual Review of Condensed Matter
  Physics}\ }\textbf {\bibinfo {volume} {6}},\ \bibinfo {pages} {15} (\bibinfo
  {year} {2015})}\BibitemShut {NoStop}%
\bibitem [{\citenamefont {Altman}\ and\ \citenamefont
  {Vosk}(2015)}]{AltmanVosk}%
  \BibitemOpen
  \bibfield  {author} {\bibinfo {author} {\bibfnamefont {E.}~\bibnamefont
  {Altman}}\ and\ \bibinfo {author} {\bibfnamefont {R.}~\bibnamefont {Vosk}},\
  }\href {\doibase 10.1146/annurev-conmatphys-031214-014701} {\bibfield
  {journal} {\bibinfo  {journal} {Annual Review of Condensed Matter Physics}\
  }\textbf {\bibinfo {volume} {6}},\ \bibinfo {pages} {383} (\bibinfo {year}
  {2015})}\BibitemShut {NoStop}%
\bibitem [{\citenamefont {Bauer}\ and\ \citenamefont {Nayak}(2013)}]{Bauer13}%
  \BibitemOpen
  \bibfield  {author} {\bibinfo {author} {\bibfnamefont {B.}~\bibnamefont
  {Bauer}}\ and\ \bibinfo {author} {\bibfnamefont {C.}~\bibnamefont {Nayak}},\
  }\href {http://stacks.iop.org/1742-5468/2013/i=09/a=P09005} {\bibfield
  {journal} {\bibinfo  {journal} {Journal of Statistical Mechanics: Theory and
  Experiment}\ }\textbf {\bibinfo {volume} {2013}},\ \bibinfo {pages} {P09005}
  (\bibinfo {year} {2013})}\BibitemShut {NoStop}%
\bibitem [{\citenamefont {Pekker}\ and\ \citenamefont
  {Clark}(2017)}]{PekkerClark}%
  \BibitemOpen
  \bibfield  {author} {\bibinfo {author} {\bibfnamefont {D.}~\bibnamefont
  {Pekker}}\ and\ \bibinfo {author} {\bibfnamefont {B.~K.}\ \bibnamefont
  {Clark}},\ }\href {\doibase 10.1103/PhysRevB.95.035116} {\bibfield  {journal}
  {\bibinfo  {journal} {Phys. Rev. B}\ }\textbf {\bibinfo {volume} {95}},\
  \bibinfo {pages} {035116} (\bibinfo {year} {2017})}\BibitemShut {NoStop}%
\bibitem [{\citenamefont {Chandran}\ \emph {et~al.}(2015)\citenamefont
  {Chandran}, \citenamefont {Carrasquilla}, \citenamefont {Kim}, \citenamefont
  {Abanin},\ and\ \citenamefont {Vidal}}]{ChandranSpectral}%
  \BibitemOpen
  \bibfield  {author} {\bibinfo {author} {\bibfnamefont {A.}~\bibnamefont
  {Chandran}}, \bibinfo {author} {\bibfnamefont {J.}~\bibnamefont
  {Carrasquilla}}, \bibinfo {author} {\bibfnamefont {I.~H.}\ \bibnamefont
  {Kim}}, \bibinfo {author} {\bibfnamefont {D.~A.}\ \bibnamefont {Abanin}}, \
  and\ \bibinfo {author} {\bibfnamefont {G.}~\bibnamefont {Vidal}},\ }\href
  {\doibase 10.1103/PhysRevB.92.024201} {\bibfield  {journal} {\bibinfo
  {journal} {Phys. Rev. B}\ }\textbf {\bibinfo {volume} {92}},\ \bibinfo
  {pages} {024201} (\bibinfo {year} {2015})}\BibitemShut {NoStop}%
\bibitem [{\citenamefont {Huse}\ \emph {et~al.}(2013)\citenamefont {Huse},
  \citenamefont {Nandkishore}, \citenamefont {Oganesyan}, \citenamefont {Pal},\
  and\ \citenamefont {Sondhi}}]{Huse13}%
  \BibitemOpen
  \bibfield  {author} {\bibinfo {author} {\bibfnamefont {D.~A.}\ \bibnamefont
  {Huse}}, \bibinfo {author} {\bibfnamefont {R.}~\bibnamefont {Nandkishore}},
  \bibinfo {author} {\bibfnamefont {V.}~\bibnamefont {Oganesyan}}, \bibinfo
  {author} {\bibfnamefont {A.}~\bibnamefont {Pal}}, \ and\ \bibinfo {author}
  {\bibfnamefont {S.~L.}\ \bibnamefont {Sondhi}},\ }\href
  {http://link.aps.org/doi/10.1103/PhysRevB.88.014206} {\bibfield  {journal}
  {\bibinfo  {journal} {Phys. Rev. B}\ }\textbf {\bibinfo {volume} {88}},\
  \bibinfo {pages} {014206} (\bibinfo {year} {2013})}\BibitemShut {NoStop}%
\bibitem [{\citenamefont {Pekker}\ \emph {et~al.}(2014)\citenamefont {Pekker},
  \citenamefont {Refael}, \citenamefont {Altman}, \citenamefont {Demler},\ and\
  \citenamefont {Oganesyan}}]{PekkerHilbertGlass}%
  \BibitemOpen
  \bibfield  {author} {\bibinfo {author} {\bibfnamefont {D.}~\bibnamefont
  {Pekker}}, \bibinfo {author} {\bibfnamefont {G.}~\bibnamefont {Refael}},
  \bibinfo {author} {\bibfnamefont {E.}~\bibnamefont {Altman}}, \bibinfo
  {author} {\bibfnamefont {E.}~\bibnamefont {Demler}}, \ and\ \bibinfo {author}
  {\bibfnamefont {V.}~\bibnamefont {Oganesyan}},\ }\href
  {http://link.aps.org/doi/10.1103/PhysRevX.4.011052} {\bibfield  {journal}
  {\bibinfo  {journal} {Phys. Rev. X}\ }\textbf {\bibinfo {volume} {4}},\
  \bibinfo {pages} {011052} (\bibinfo {year} {2014})}\BibitemShut {NoStop}%
\bibitem [{\citenamefont {{Parameswaran}}\ \emph {et~al.}(2016)\citenamefont
  {{Parameswaran}}, \citenamefont {{Potter}},\ and\ \citenamefont
  {{Vasseur}}}]{EigenstateReview}%
  \BibitemOpen
  \bibfield  {author} {\bibinfo {author} {\bibfnamefont {S.~A.}\ \bibnamefont
  {{Parameswaran}}}, \bibinfo {author} {\bibfnamefont {A.~C.}\ \bibnamefont
  {{Potter}}}, \ and\ \bibinfo {author} {\bibfnamefont {R.}~\bibnamefont
  {{Vasseur}}},\ }\href@noop {} {\bibfield  {journal} {\bibinfo  {journal}
  {ArXiv e-prints}\ } (\bibinfo {year} {2016})},\ \Eprint
  {http://arxiv.org/abs/1610.03078} {arXiv:1610.03078 [cond-mat.dis-nn]}
  \BibitemShut {NoStop}%
\bibitem [{\citenamefont {Oganesyan}\ and\ \citenamefont
  {Huse}(2007{\natexlab{b}})}]{OganesyanMBL}%
  \BibitemOpen
  \bibfield  {author} {\bibinfo {author} {\bibfnamefont {V.}~\bibnamefont
  {Oganesyan}}\ and\ \bibinfo {author} {\bibfnamefont {D.~A.}\ \bibnamefont
  {Huse}},\ }\href {\doibase 10.1103/PhysRevB.75.155111} {\bibfield  {journal}
  {\bibinfo  {journal} {Phys. Rev. B}\ }\textbf {\bibinfo {volume} {75}},\
  \bibinfo {pages} {155111} (\bibinfo {year} {2007}{\natexlab{b}})}\BibitemShut
  {NoStop}%
\bibitem [{\citenamefont {Kj\"all}\ \emph {et~al.}(2014)\citenamefont
  {Kj\"all}, \citenamefont {Bardarson},\ and\ \citenamefont
  {Pollmann}}]{KjallMBL}%
  \BibitemOpen
  \bibfield  {author} {\bibinfo {author} {\bibfnamefont {J.~A.}\ \bibnamefont
  {Kj\"all}}, \bibinfo {author} {\bibfnamefont {J.~H.}\ \bibnamefont
  {Bardarson}}, \ and\ \bibinfo {author} {\bibfnamefont {F.}~\bibnamefont
  {Pollmann}},\ }\href {\doibase 10.1103/PhysRevLett.113.107204} {\bibfield
  {journal} {\bibinfo  {journal} {Phys. Rev. Lett.}\ }\textbf {\bibinfo
  {volume} {113}},\ \bibinfo {pages} {107204} (\bibinfo {year}
  {2014})}\BibitemShut {NoStop}%
\bibitem [{\citenamefont {Vosk}\ \emph {et~al.}(2015)\citenamefont {Vosk},
  \citenamefont {Huse},\ and\ \citenamefont {Altman}}]{VoskMBL}%
  \BibitemOpen
  \bibfield  {author} {\bibinfo {author} {\bibfnamefont {R.}~\bibnamefont
  {Vosk}}, \bibinfo {author} {\bibfnamefont {D.~A.}\ \bibnamefont {Huse}}, \
  and\ \bibinfo {author} {\bibfnamefont {E.}~\bibnamefont {Altman}},\ }\href
  {\doibase 10.1103/PhysRevX.5.031032} {\bibfield  {journal} {\bibinfo
  {journal} {Phys. Rev. X}\ }\textbf {\bibinfo {volume} {5}},\ \bibinfo {pages}
  {031032} (\bibinfo {year} {2015})}\BibitemShut {NoStop}%
\bibitem [{\citenamefont {Potter}\ \emph {et~al.}(2015)\citenamefont {Potter},
  \citenamefont {Vasseur},\ and\ \citenamefont {Parameswaran}}]{PotterMBL}%
  \BibitemOpen
  \bibfield  {author} {\bibinfo {author} {\bibfnamefont {A.~C.}\ \bibnamefont
  {Potter}}, \bibinfo {author} {\bibfnamefont {R.}~\bibnamefont {Vasseur}}, \
  and\ \bibinfo {author} {\bibfnamefont {S.~A.}\ \bibnamefont {Parameswaran}},\
  }\href {\doibase 10.1103/PhysRevX.5.031033} {\bibfield  {journal} {\bibinfo
  {journal} {Phys. Rev. X}\ }\textbf {\bibinfo {volume} {5}},\ \bibinfo {pages}
  {031033} (\bibinfo {year} {2015})}\BibitemShut {NoStop}%
\bibitem [{\citenamefont {Devakul}\ and\ \citenamefont
  {Singh}(2015{\natexlab{a}})}]{DevakulMBL}%
  \BibitemOpen
  \bibfield  {author} {\bibinfo {author} {\bibfnamefont {T.}~\bibnamefont
  {Devakul}}\ and\ \bibinfo {author} {\bibfnamefont {R.~R.~P.}\ \bibnamefont
  {Singh}},\ }\href {\doibase 10.1103/PhysRevLett.115.187201} {\bibfield
  {journal} {\bibinfo  {journal} {Phys. Rev. Lett.}\ }\textbf {\bibinfo
  {volume} {115}},\ \bibinfo {pages} {187201} (\bibinfo {year}
  {2015}{\natexlab{a}})}\BibitemShut {NoStop}%
\bibitem [{\citenamefont {Luitz}\ \emph
  {et~al.}(2015{\natexlab{a}})\citenamefont {Luitz}, \citenamefont
  {Laflorencie},\ and\ \citenamefont {Alet}}]{LuitzMBL}%
  \BibitemOpen
  \bibfield  {author} {\bibinfo {author} {\bibfnamefont {D.~J.}\ \bibnamefont
  {Luitz}}, \bibinfo {author} {\bibfnamefont {N.}~\bibnamefont {Laflorencie}},
  \ and\ \bibinfo {author} {\bibfnamefont {F.}~\bibnamefont {Alet}},\ }\href
  {\doibase 10.1103/PhysRevB.91.081103} {\bibfield  {journal} {\bibinfo
  {journal} {Phys. Rev. B}\ }\textbf {\bibinfo {volume} {91}},\ \bibinfo
  {pages} {081103} (\bibinfo {year} {2015}{\natexlab{a}})}\BibitemShut
  {NoStop}%
\bibitem [{\citenamefont {Zhang}\ \emph
  {et~al.}(2016{\natexlab{a}})\citenamefont {Zhang}, \citenamefont {Zhao},
  \citenamefont {Devakul},\ and\ \citenamefont {Huse}}]{ZhangMBL}%
  \BibitemOpen
  \bibfield  {author} {\bibinfo {author} {\bibfnamefont {L.}~\bibnamefont
  {Zhang}}, \bibinfo {author} {\bibfnamefont {B.}~\bibnamefont {Zhao}},
  \bibinfo {author} {\bibfnamefont {T.}~\bibnamefont {Devakul}}, \ and\
  \bibinfo {author} {\bibfnamefont {D.~A.}\ \bibnamefont {Huse}},\ }\href
  {\doibase 10.1103/PhysRevB.93.224201} {\bibfield  {journal} {\bibinfo
  {journal} {Phys. Rev. B}\ }\textbf {\bibinfo {volume} {93}},\ \bibinfo
  {pages} {224201} (\bibinfo {year} {2016}{\natexlab{a}})}\BibitemShut
  {NoStop}%
\bibitem [{\citenamefont {Serbyn}\ and\ \citenamefont
  {Moore}(2016)}]{SerbynMBL}%
  \BibitemOpen
  \bibfield  {author} {\bibinfo {author} {\bibfnamefont {M.}~\bibnamefont
  {Serbyn}}\ and\ \bibinfo {author} {\bibfnamefont {J.~E.}\ \bibnamefont
  {Moore}},\ }\href {\doibase 10.1103/PhysRevB.93.041424} {\bibfield  {journal}
  {\bibinfo  {journal} {Phys. Rev. B}\ }\textbf {\bibinfo {volume} {93}},\
  \bibinfo {pages} {041424} (\bibinfo {year} {2016})}\BibitemShut {NoStop}%
\bibitem [{\citenamefont {Zhang}\ \emph
  {et~al.}(2016{\natexlab{b}})\citenamefont {Zhang}, \citenamefont {Khemani},\
  and\ \citenamefont {Huse}}]{ZhangMBL2}%
  \BibitemOpen
  \bibfield  {author} {\bibinfo {author} {\bibfnamefont {L.}~\bibnamefont
  {Zhang}}, \bibinfo {author} {\bibfnamefont {V.}~\bibnamefont {Khemani}}, \
  and\ \bibinfo {author} {\bibfnamefont {D.~A.}\ \bibnamefont {Huse}},\ }\href
  {\doibase 10.1103/PhysRevB.94.224202} {\bibfield  {journal} {\bibinfo
  {journal} {Phys. Rev. B}\ }\textbf {\bibinfo {volume} {94}},\ \bibinfo
  {pages} {224202} (\bibinfo {year} {2016}{\natexlab{b}})}\BibitemShut
  {NoStop}%
\bibitem [{\citenamefont {V.~Khemani}\ and\ \citenamefont
  {Huse}(2017)}]{KhemaniMBL}%
  \BibitemOpen
  \bibfield  {author} {\bibinfo {author} {\bibfnamefont {D.~N.~S.}\
  \bibnamefont {V.~Khemani}}\ and\ \bibinfo {author} {\bibfnamefont {D.~A.}\
  \bibnamefont {Huse}},\ }\href@noop {} {\bibfield  {journal} {\bibinfo
  {journal} {arXiv:102.03932}\ } (\bibinfo {year} {2017})}\BibitemShut
  {NoStop}%
\bibitem [{\citenamefont {Devakul}\ and\ \citenamefont
  {Singh}(2015{\natexlab{b}})}]{Devakul15}%
  \BibitemOpen
  \bibfield  {author} {\bibinfo {author} {\bibfnamefont {T.}~\bibnamefont
  {Devakul}}\ and\ \bibinfo {author} {\bibfnamefont {R.~R.~P.}\ \bibnamefont
  {Singh}},\ }\href {\doibase 10.1103/PhysRevLett.115.187201} {\bibfield
  {journal} {\bibinfo  {journal} {Phys. Rev. Lett.}\ }\textbf {\bibinfo
  {volume} {115}},\ \bibinfo {pages} {187201} (\bibinfo {year}
  {2015}{\natexlab{b}})}\BibitemShut {NoStop}%
\bibitem [{\citenamefont {Luitz}\ \emph
  {et~al.}(2015{\natexlab{b}})\citenamefont {Luitz}, \citenamefont
  {Laflorencie},\ and\ \citenamefont {Alet}}]{Luitz15}%
  \BibitemOpen
  \bibfield  {author} {\bibinfo {author} {\bibfnamefont {D.~J.}\ \bibnamefont
  {Luitz}}, \bibinfo {author} {\bibfnamefont {N.}~\bibnamefont {Laflorencie}},
  \ and\ \bibinfo {author} {\bibfnamefont {F.}~\bibnamefont {Alet}},\ }\href
  {\doibase 10.1103/PhysRevB.91.081103} {\bibfield  {journal} {\bibinfo
  {journal} {Phys. Rev. B}\ }\textbf {\bibinfo {volume} {91}},\ \bibinfo
  {pages} {081103} (\bibinfo {year} {2015}{\natexlab{b}})}\BibitemShut
  {NoStop}%
\bibitem [{\citenamefont {{Khemani}}\ \emph {et~al.}(2016)\citenamefont
  {{Khemani}}, \citenamefont {{Lim}}, \citenamefont {{Sheng}},\ and\
  \citenamefont {{Huse}}}]{KhemaniCP}%
  \BibitemOpen
  \bibfield  {author} {\bibinfo {author} {\bibfnamefont {V.}~\bibnamefont
  {{Khemani}}}, \bibinfo {author} {\bibfnamefont {S.~P.}\ \bibnamefont
  {{Lim}}}, \bibinfo {author} {\bibfnamefont {D.~N.}\ \bibnamefont {{Sheng}}},
  \ and\ \bibinfo {author} {\bibfnamefont {D.~A.}\ \bibnamefont {{Huse}}},\
  }\href@noop {} {\bibfield  {journal} {\bibinfo  {journal} {ArXiv e-prints}\ }
  (\bibinfo {year} {2016})},\ \Eprint {http://arxiv.org/abs/1607.05756}
  {arXiv:1607.05756 [cond-mat.dis-nn]} \BibitemShut {NoStop}%
\bibitem [{\citenamefont {{Grover}}(2014)}]{GroverCP}%
  \BibitemOpen
  \bibfield  {author} {\bibinfo {author} {\bibfnamefont {T.}~\bibnamefont
  {{Grover}}},\ }\href@noop {} {\bibfield  {journal} {\bibinfo  {journal}
  {ArXiv e-prints}\ } (\bibinfo {year} {2014})},\ \Eprint
  {http://arxiv.org/abs/1405.1471} {arXiv:1405.1471 [cond-mat.dis-nn]}
  \BibitemShut {NoStop}%
\bibitem [{\citenamefont {Pollmann}\ \emph {et~al.}(2016)\citenamefont
  {Pollmann}, \citenamefont {Khemani}, \citenamefont {Cirac},\ and\
  \citenamefont {Sondhi}}]{PollmannVUMPO}%
  \BibitemOpen
  \bibfield  {author} {\bibinfo {author} {\bibfnamefont {F.}~\bibnamefont
  {Pollmann}}, \bibinfo {author} {\bibfnamefont {V.}~\bibnamefont {Khemani}},
  \bibinfo {author} {\bibfnamefont {J.~I.}\ \bibnamefont {Cirac}}, \ and\
  \bibinfo {author} {\bibfnamefont {S.~L.}\ \bibnamefont {Sondhi}},\ }\href
  {\doibase 10.1103/PhysRevB.94.041116} {\bibfield  {journal} {\bibinfo
  {journal} {Phys. Rev. B}\ }\textbf {\bibinfo {volume} {94}},\ \bibinfo
  {pages} {041116} (\bibinfo {year} {2016})}\BibitemShut {NoStop}%
\bibitem [{\citenamefont {{Wahl}}\ \emph {et~al.}(2016)\citenamefont {{Wahl}},
  \citenamefont {{Pal}},\ and\ \citenamefont {{Simon}}}]{PalSimon}%
  \BibitemOpen
  \bibfield  {author} {\bibinfo {author} {\bibfnamefont {T.~B.}\ \bibnamefont
  {{Wahl}}}, \bibinfo {author} {\bibfnamefont {A.}~\bibnamefont {{Pal}}}, \
  and\ \bibinfo {author} {\bibfnamefont {S.~H.}\ \bibnamefont {{Simon}}},\
  }\href@noop {} {\bibfield  {journal} {\bibinfo  {journal} {ArXiv e-prints}\ }
  (\bibinfo {year} {2016})},\ \Eprint {http://arxiv.org/abs/1609.01552}
  {arXiv:1609.01552 [cond-mat.dis-nn]} \BibitemShut {NoStop}%
\bibitem [{\citenamefont {Khemani}\ \emph {et~al.}(2016)\citenamefont
  {Khemani}, \citenamefont {Pollmann},\ and\ \citenamefont
  {Sondhi}}]{KhemaniDMRGX}%
  \BibitemOpen
  \bibfield  {author} {\bibinfo {author} {\bibfnamefont {V.}~\bibnamefont
  {Khemani}}, \bibinfo {author} {\bibfnamefont {F.}~\bibnamefont {Pollmann}}, \
  and\ \bibinfo {author} {\bibfnamefont {S.~L.}\ \bibnamefont {Sondhi}},\
  }\href {http://link.aps.org/doi/10.1103/PhysRevLett.116.247204} {\bibfield
  {journal} {\bibinfo  {journal} {Phys. Rev. Lett.}\ }\textbf {\bibinfo
  {volume} {116}},\ \bibinfo {pages} {247204} (\bibinfo {year}
  {2016})}\BibitemShut {NoStop}%
\bibitem [{\citenamefont {Yu}\ \emph {et~al.}(2017)\citenamefont {Yu},
  \citenamefont {Pekker},\ and\ \citenamefont {Clark}}]{YuPekkerClark}%
  \BibitemOpen
  \bibfield  {author} {\bibinfo {author} {\bibfnamefont {X.}~\bibnamefont
  {Yu}}, \bibinfo {author} {\bibfnamefont {D.}~\bibnamefont {Pekker}}, \ and\
  \bibinfo {author} {\bibfnamefont {B.~K.}\ \bibnamefont {Clark}},\ }\href
  {\doibase 10.1103/PhysRevLett.118.017201} {\bibfield  {journal} {\bibinfo
  {journal} {Phys. Rev. Lett.}\ }\textbf {\bibinfo {volume} {118}},\ \bibinfo
  {pages} {017201} (\bibinfo {year} {2017})}\BibitemShut {NoStop}%
\bibitem [{\citenamefont {Lim}\ and\ \citenamefont {Sheng}(2016)}]{LimSheng}%
  \BibitemOpen
  \bibfield  {author} {\bibinfo {author} {\bibfnamefont {S.~P.}\ \bibnamefont
  {Lim}}\ and\ \bibinfo {author} {\bibfnamefont {D.~N.}\ \bibnamefont
  {Sheng}},\ }\href {\doibase 10.1103/PhysRevB.94.045111} {\bibfield  {journal}
  {\bibinfo  {journal} {Phys. Rev. B}\ }\textbf {\bibinfo {volume} {94}},\
  \bibinfo {pages} {045111} (\bibinfo {year} {2016})}\BibitemShut {NoStop}%
\bibitem [{\citenamefont {Serbyn}\ \emph {et~al.}(2016)\citenamefont {Serbyn},
  \citenamefont {Michailidis}, \citenamefont {Abanin},\ and\ \citenamefont
  {Papi\ifmmode~\acute{c}\else \'{c}\fi{}}}]{SerbynPowerlaw}%
  \BibitemOpen
  \bibfield  {author} {\bibinfo {author} {\bibfnamefont {M.}~\bibnamefont
  {Serbyn}}, \bibinfo {author} {\bibfnamefont {A.~A.}\ \bibnamefont
  {Michailidis}}, \bibinfo {author} {\bibfnamefont {D.~A.}\ \bibnamefont
  {Abanin}}, \ and\ \bibinfo {author} {\bibfnamefont {Z.}~\bibnamefont
  {Papi\ifmmode~\acute{c}\else \'{c}\fi{}}},\ }\href {\doibase
  10.1103/PhysRevLett.117.160601} {\bibfield  {journal} {\bibinfo  {journal}
  {Phys. Rev. Lett.}\ }\textbf {\bibinfo {volume} {117}},\ \bibinfo {pages}
  {160601} (\bibinfo {year} {2016})}\BibitemShut {NoStop}%
\bibitem [{\citenamefont {Kennes}\ and\ \citenamefont
  {Karrasch}(2016)}]{Dante}%
  \BibitemOpen
  \bibfield  {author} {\bibinfo {author} {\bibfnamefont {D.~M.}\ \bibnamefont
  {Kennes}}\ and\ \bibinfo {author} {\bibfnamefont {C.}~\bibnamefont
  {Karrasch}},\ }\href {\doibase 10.1103/PhysRevB.93.245129} {\bibfield
  {journal} {\bibinfo  {journal} {Phys. Rev. B}\ }\textbf {\bibinfo {volume}
  {93}},\ \bibinfo {pages} {245129} (\bibinfo {year} {2016})}\BibitemShut
  {NoStop}%
\bibitem [{\citenamefont {Huse}\ \emph {et~al.}(2014)\citenamefont {Huse},
  \citenamefont {Nandkishore},\ and\ \citenamefont {Oganesyan}}]{Huse14}%
  \BibitemOpen
  \bibfield  {author} {\bibinfo {author} {\bibfnamefont {D.~A.}\ \bibnamefont
  {Huse}}, \bibinfo {author} {\bibfnamefont {R.}~\bibnamefont {Nandkishore}}, \
  and\ \bibinfo {author} {\bibfnamefont {V.}~\bibnamefont {Oganesyan}},\ }\href
  {\doibase 10.1103/PhysRevB.90.174202} {\bibfield  {journal} {\bibinfo
  {journal} {Phys. Rev. B}\ }\textbf {\bibinfo {volume} {90}},\ \bibinfo
  {pages} {174202} (\bibinfo {year} {2014})}\BibitemShut {NoStop}%
\bibitem [{\citenamefont {Serbyn}\ \emph {et~al.}(2013)\citenamefont {Serbyn},
  \citenamefont {Papi\ifmmode~\acute{c}\else \'{c}\fi{}},\ and\ \citenamefont
  {Abanin}}]{Serbyn13cons}%
  \BibitemOpen
  \bibfield  {author} {\bibinfo {author} {\bibfnamefont {M.}~\bibnamefont
  {Serbyn}}, \bibinfo {author} {\bibfnamefont {Z.}~\bibnamefont
  {Papi\ifmmode~\acute{c}\else \'{c}\fi{}}}, \ and\ \bibinfo {author}
  {\bibfnamefont {D.~A.}\ \bibnamefont {Abanin}},\ }\href
  {http://link.aps.org/doi/10.1103/PhysRevLett.111.127201} {\bibfield
  {journal} {\bibinfo  {journal} {Phys. Rev. Lett.}\ }\textbf {\bibinfo
  {volume} {111}},\ \bibinfo {pages} {127201} (\bibinfo {year}
  {2013})}\BibitemShut {NoStop}%
\bibitem [{\citenamefont {Vidal}(2003)}]{VidalCanonical}%
  \BibitemOpen
  \bibfield  {author} {\bibinfo {author} {\bibfnamefont {G.}~\bibnamefont
  {Vidal}},\ }\href {\doibase 10.1103/PhysRevLett.91.147902} {\bibfield
  {journal} {\bibinfo  {journal} {Phys. Rev. Lett.}\ }\textbf {\bibinfo
  {volume} {91}},\ \bibinfo {pages} {147902} (\bibinfo {year}
  {2003})}\BibitemShut {NoStop}%
\bibitem [{\citenamefont {{Kappus M.}}\ and\ \citenamefont {{Wegner
  F.}}(1981)}]{Kappus1981}%
  \BibitemOpen
  \bibfield  {author} {\bibinfo {author} {\bibnamefont {{Kappus M.}}}\ and\
  \bibinfo {author} {\bibnamefont {{Wegner F.}}},\ }\href {\doibase
  http://dx.doi.org/10.1007/BF01294272} {\bibfield  {journal} {\bibinfo
  {journal} {Zeitschrift f{\"u}r Physik B Condensed Matter}\ }\textbf {\bibinfo
  {volume} {45}},\ \bibinfo {pages} {15} (\bibinfo {year} {1981})}\BibitemShut
  {NoStop}%
\bibitem [{Note2()}]{Note2}%
  \BibitemOpen
  \bibinfo {note} {They are obviously local if one takes the definition of an
  MBL system to be the existence of a finite depth unitary transformation that
  diagonalizes the Hamiltonian.}\BibitemShut {Stop}%
\bibitem [{Note3()}]{Note3}%
  \BibitemOpen
  \bibinfo {note} {The argument in Ref~\cite {SerbynPowerlaw} relies on the
  l-bit operators flipping many spins within a radius $r$, but in a
  noninteracting model the l-bit operators only flip two.}\BibitemShut {Stop}%
\bibitem [{\citenamefont {Eisler}(2009)}]{Peschel}%
  \BibitemOpen
  \bibfield  {author} {\bibinfo {author} {\bibfnamefont {I.}~\bibnamefont
  {Eisler}},\ }\href {http://stacks.iop.org/1751-8121/42/i=50/a=504003}
  {\bibfield  {journal} {\bibinfo  {journal} {Journal of Physics A:
  Mathematical and Theoretical}\ }\textbf {\bibinfo {volume} {42}},\ \bibinfo
  {pages} {504003} (\bibinfo {year} {2009})}\BibitemShut {NoStop}%
\bibitem [{Note4()}]{Note4}%
  \BibitemOpen
  \bibinfo {note} {Finding the largest $k$ eigenvalues of $\rho _\protect \text
  {A}$ involves finding the $k$ lowest total eigenenergies given all the single
  particle energies, which is a standard computational problem.}\BibitemShut
  {Stop}%
\bibitem [{\citenamefont {Fishman}\ and\ \citenamefont
  {White}(2015)}]{WhiteCompression}%
  \BibitemOpen
  \bibfield  {author} {\bibinfo {author} {\bibfnamefont {M.~T.}\ \bibnamefont
  {Fishman}}\ and\ \bibinfo {author} {\bibfnamefont {S.~R.}\ \bibnamefont
  {White}},\ }\href {\doibase 10.1103/PhysRevB.92.075132} {\bibfield  {journal}
  {\bibinfo  {journal} {Phys. Rev. B}\ }\textbf {\bibinfo {volume} {92}},\
  \bibinfo {pages} {075132} (\bibinfo {year} {2015})}\BibitemShut {NoStop}%
\bibitem [{\citenamefont {Silvi}\ \emph {et~al.}(2013)\citenamefont {Silvi},
  \citenamefont {Rossini}, \citenamefont {Fazio}, \citenamefont {Santoro},\
  and\ \citenamefont {Giovannetti}}]{Silvi}%
  \BibitemOpen
  \bibfield  {author} {\bibinfo {author} {\bibfnamefont {P.}~\bibnamefont
  {Silvi}}, \bibinfo {author} {\bibfnamefont {D.}~\bibnamefont {Rossini}},
  \bibinfo {author} {\bibfnamefont {R.}~\bibnamefont {Fazio}}, \bibinfo
  {author} {\bibfnamefont {G.~E.}\ \bibnamefont {Santoro}}, \ and\ \bibinfo
  {author} {\bibfnamefont {V.}~\bibnamefont {Giovannetti}},\ }\href {\doibase
  10.1142/S021797921345029X} {\bibfield  {journal} {\bibinfo  {journal}
  {International Journal of Modern Physics B}\ }\textbf {\bibinfo {volume}
  {27}},\ \bibinfo {pages} {1345029} (\bibinfo {year} {2013})},\ \Eprint
  {http://arxiv.org/abs/http://www.worldscientific.com/doi/pdf/10.1142/S021797921345029X}
  {http://www.worldscientific.com/doi/pdf/10.1142/S021797921345029X}
  \BibitemShut {NoStop}%
\bibitem [{\citenamefont {Katsura}\ and\ \citenamefont
  {Maruyama}(2010)}]{Katsura}%
  \BibitemOpen
  \bibfield  {author} {\bibinfo {author} {\bibfnamefont {H.}~\bibnamefont
  {Katsura}}\ and\ \bibinfo {author} {\bibfnamefont {I.}~\bibnamefont
  {Maruyama}},\ }\href {http://stacks.iop.org/1751-8121/43/i=17/a=175003}
  {\bibfield  {journal} {\bibinfo  {journal} {Journal of Physics A:
  Mathematical and Theoretical}\ }\textbf {\bibinfo {volume} {43}},\ \bibinfo
  {pages} {175003} (\bibinfo {year} {2010})}\BibitemShut {NoStop}%
\bibitem [{\citenamefont {Murg}\ \emph {et~al.}(2012)\citenamefont {Murg},
  \citenamefont {Korepin},\ and\ \citenamefont {Verstraete}}]{Murg}%
  \BibitemOpen
  \bibfield  {author} {\bibinfo {author} {\bibfnamefont {V.}~\bibnamefont
  {Murg}}, \bibinfo {author} {\bibfnamefont {V.~E.}\ \bibnamefont {Korepin}}, \
  and\ \bibinfo {author} {\bibfnamefont {F.}~\bibnamefont {Verstraete}},\
  }\href {\doibase 10.1103/PhysRevB.86.045125} {\bibfield  {journal} {\bibinfo
  {journal} {Phys. Rev. B}\ }\textbf {\bibinfo {volume} {86}},\ \bibinfo
  {pages} {045125} (\bibinfo {year} {2012})}\BibitemShut {NoStop}%
\bibitem [{Note5()}]{Note5}%
  \BibitemOpen
  \bibinfo {note} {Note that it is not that the algorithm of Ref~\cite
  {WhiteCompression} is failing at low $W$. Similar curves were obtained for
  smaller $L$ accessible by ED, and a standard MPS compression of the ED state
  yielded similar results.}\BibitemShut {Stop}%
\bibitem [{\citenamefont {Aubry}\ and\ \citenamefont
  {Andre}(1980)}]{aubryandre}%
  \BibitemOpen
  \bibfield  {author} {\bibinfo {author} {\bibfnamefont {S.}~\bibnamefont
  {Aubry}}\ and\ \bibinfo {author} {\bibfnamefont {G.}~\bibnamefont {Andre}},\
  }\href@noop {} {\bibfield  {journal} {\bibinfo  {journal} {Ann. Isr. Phys.
  Soc.}\ }\textbf {\bibinfo {volume} {3}},\ \bibinfo {pages} {133} (\bibinfo
  {year} {1980})}\BibitemShut {NoStop}%
\end{thebibliography}%

\begin{appendices}
    \section{Locality of l-bit operators}
    \label{sec:lbitlocality}
        We have a Jordan Wigner transformed spin chain, diagonalized by a set of fermionic operators 
        \begin{equation}
            \mathcal{H} = \sum_\alpha \epsilon_\alpha a^\dagger_\alpha a_\alpha
            \label{}
        \end{equation}
        where each $a^\dagger_\alpha = \sum_i u_\alpha(i) c^\dagger_i$, and in the spin language, $c^\dagger_i = \prod_{j<i} {(-1)}^{(S^z_j+\frac{1}{2})} S^{+}_i$, and similarly for the conjugates.
        We want to show that the bosonic raising and lowering operators $\tau^+_\alpha$ and $\tau^-_\alpha$ are also localized, given that the fermionic ones $a^\dagger_\alpha$ and $a_\alpha$ are in terms of the physical $c^\dagger_i$,$c_i$ operators.

        The corresponding bosonic raising operator for $\tau^z_\alpha = 2a^\dagger_\alpha a_\alpha - 1$ is then defined by an inverse Jordan Wigner transformation
        \begin{eqnarray}
            \tau^+_\alpha &=&  \prod_{\beta =1}^{\alpha-1} {(-1)}^{a^\dagger_\beta a_\beta} a^\dagger_\alpha\\
            &=& {(-1)}^{\sum_{\beta=1}^{\alpha-1} a^\dagger_\beta a_\beta} \sum_{i=1}^{L}u_\alpha(i) {(-1)}^{\sum_{j=1}^{i-1} c^\dagger_j c_j}S^+_i
            \label{}
        \end{eqnarray}
        For simplicity, we will assume that $u_\alpha(i)$ is strictly localized within a range $\xi$, and is therefore only nonzero if $|\alpha - i| \leq \xi$ (in reality, there will be an exponential decay with lengthscale $\xi$, and our argument can be generalized for some error threshold).
        We can then write $\tau^+_\alpha$ as
        \begin{eqnarray}
            \label{eq:factoredchain}
            \tau^+_\alpha &=& {(-1)}^{\sum_{\beta=1}^{\alpha-1} a_\beta^\dagger a_\beta} {(-1)}^{\sum_{j=1}^{\alpha-\xi-1} c_j^\dagger c_j}\nonumber \\
            && \sum_{i=\alpha-\xi}^{\alpha+\xi-1}u_\alpha(i) {(-1)}^{\sum_{j=\alpha-\xi}^{i-1} c^\dagger_j c_j}S^+_i
        \end{eqnarray}

        Expressing the first exponent in Equation~\ref{eq:factoredchain} as
        \begin{eqnarray}
            \sum_{\beta =1}^{\alpha-1} a^\dagger_\beta a_\beta 
            &=&  \sum_{j,k=1}^{L} \sum_{\beta=1}^{\alpha-1} u_\beta(j)u^*_\beta(k)c^\dagger_j c_k 
            \label{}
        \end{eqnarray}
        we see that if $j < \alpha - \xi$ or $k < \alpha - \xi$, then the sum over $\beta$ contains all non-zero elements of $u_\beta(i) u^*_\beta(j)$, resulting in a dirac delta $\delta_{i,j}$ by completeness.
        This reduces the expression to
        \begin{eqnarray}
            \sum_{\beta=1}^{\alpha-1} a^\dagger_\beta a_\beta 
            &=& \sum_{j=1}^{\alpha - \xi-1} c_j^\dagger c_j + \nonumber\\
            && \sum_{j,k=\alpha-\xi}^{\alpha+\xi-1} \sum_{\beta=1}^{\alpha-1} u_\beta(j)u^*_\beta(k)c^\dagger_j c_k 
            \label{eq:twosums}
        \end{eqnarray}
        where we have restricted the summation indices $j,k \leq \alpha+\xi-1$, since $\beta \leq \alpha-1$ and we have assumed $u_\beta(i)$ is zero for $i>\beta+\xi$.

        The two sums in Eq~\ref{eq:twosums} commute, and therefore the $c^\dagger_j c_j$ sum can be canceled out with the chain in Eq~\ref{eq:factoredchain}, leaving
        \begin{eqnarray}
            \tau^+_\alpha &=&  {(-1)}^{\sum_{j,k=\alpha-\xi}^{\alpha+\xi-1} \sum_{\beta=1}^{\alpha-1} u_\beta(j)u^*_\beta(k)c^\dagger_j c_k 
}\\
&&\sum_{i=\alpha-\xi}^{\alpha+\xi}u_\alpha(i) {(-1)}^{\sum_{j=\alpha-\xi}^{i-1} c^\dagger_j c_j}S^+_i
        \end{eqnarray}
        which is strictly localized within $\alpha \pm \xi$.

        \section{MPO Representation of l-bit operators}
        \label{sec:lbitMPO}

        The $a^\dagger_\alpha a_\alpha$ operators can be constructed as an MPO by hand.
        The methodology is similar to that used to construct the MPO representation of the Hamiltonian~\cite{Schollwock2011}.
        We use the internal dimensions $\{s_i\}$ to keep track of the different terms in the expansion of $a^\dagger_\alpha a_\alpha$.
        We can define the full operator as an MPO, acting on spin in the $S^z$ basis,
        \begin{eqnarray}
            \langle \{\sigma_i'\}|a^\dagger_\alpha a_\alpha|\{\sigma_i\}\rangle &&  \nonumber\\
            = \sum_{\{s_i\}}
            \delta_{s_i,1}\delta_{s_{L+1},4} &&\prod_{i=1}^{L} \langle \sigma_i' | M^{[i]}_{s_i,s_{i+1}} | \sigma_i \rangle
            \label{}
        \end{eqnarray}
        where each $s_i$ goes from $1$ to $4$ and $\delta_{i,j}$ is the Dirac delta.
        
        The different terms in $a^\dagger_\alpha a_\alpha = \sum_{i,j} u_\alpha(i) u^*_\alpha(j) c^\dagger_i c_j$ can be taken into account by defining $M^{[i]}_{s_i,s_{i+1}}$ appropriately:
        \begin{equation}
            M^{[i]} = 
            \begin{bmatrix}
            \mathbb{I} & u_\alpha(i)S^+ & u^*_\alpha(i)S^- & |u_\alpha(i)|^2 (S^z+1/2) \\
            0 & -2S^z & 0 & u^*_\alpha(i)S^- \\
            0 & 0 & -2S^z & u_\alpha(i)S^+ \\
            0 & 0 & 0 & \mathbb{I}
            \end{bmatrix}
            \label{}
        \end{equation}

        Thus, the operator $a^\dagger_\alpha a_\alpha$ has been expressed as an MPO of internal dimension $4$, and the expectation value in an MPS can be efficiently calculated.
        One can then simply take $\tau^z_\alpha = 2 a_\alpha^\dagger a_\alpha - 1$.

         \section{Griffiths Effects and Quasiperiodicity}
         \label{sec:qp}
%We end with an investigation of potential Griffiths-like effects in quasiperiodic systems.
Many of the difficulties in studying disordered systems using DMRG arise due to the existence of locally clean-looking patches, which are what the tail of the distribution $P(S_E)$ consist of.
An exponentially decaying tail in $P(S_E)$ will lead to the typical maximum entanglement in a system of size $L$ growing as $S_E^\text{max}\sim \log L$, resulting in the computational difficulty of DMRG grow faster than linearly.
To avoid the difficulties of these Griffiths region in studies of localization, one possibility is to examine quasiperiodic systems.

One commonly studied possibility is the Aubry-Andr\'e model, which is given by the Hamiltonian
\begin{equation}
    \mathcal{H}_\text{AA} = -\sum_i \left(c^\dagger_i c_{i+1} + \text{h.c.} \right)+ W\sum_i V(i) c_i^\dagger c_i
    \label{}
\end{equation}
where $V(i)=\cos(2\pi \omega (i + \phi))$,  $\omega$ is an irrational number which we choose to be 1 over the golden ratio, and $\phi$ is some offset.
The exact delocalization-localization critical point of this model is known to be at $W=2$, and the critical wavefunctions can be found exactly~\cite{aubryandre}.
The quasiperiodic potential does not allow for large Griffiths-like regions to exist.

Despite this, sampling chains of length $L$ (over different phase offsets), there is still a clear logarithmic growth of $S^\text{max}_E$ with $L$, corresponding to an exponentially decaying tail in $P(S_E)$, which is reminiscent of the existence of Griffiths regions!

This effect can be understood as coming from the existence of special reflection symmetry points in the quasiperiodic potential.
For example, if $\phi=0$, all the sites on the left side of $i=0$ will be in exact resonance with another site on the right side, $V(-i) = V(i)$.
Each resonant pair will then form an odd-even superposition and the many-body entanglement entropy across that cut will grow approximately as $S_E^\text{max}\sim L$.
Thus, there exist special values of $\phi$ for which the entanglement diverges across a cut.

In an actual sample, this exact resonances will not occur, only approximate resonances.
Each sample can be thought of as sampling $\phi$ $L$ times, and so one will typically be at most $\sim 1/L$ away from the exact resonance point, and sites will be detuned by $\sim 1/L$.
The effective hopping between two sites distance $x$ apart scales as $\sim e^{-x/\xi}$ for localization length $\xi$.
Therefore, for such a near resonance, only pairs of sites within range $x$ such that $e^{-x/\xi} \gtrsim 1/L$ will form superpositions.
This leads to an entanglement going as $S_E^\text{max} \sim x \sim \log L$.
Thus, a completely different mechanism in the quasiperiodic model is allowing for logarithmic growth in maximum entanglement.

To avoid a logarithmic growth of $S_E^\text{max}$, one should then avoid a quasiperiodic potential with reflection symmetry.
This is possible by a potential such as $V(i) = \cos(2\pi \omega (i+\phi)) + \sin(4\pi \omega (i+\phi))$, which does not exhibit any even reflection symmetry.
The $S_E$ distribution for this model has a very sharp cutoff and the maximum entanglement therefore does not keep growing as $\log L$.

\end{appendices}

\end{document}